\documentclass[twocolumn]{aastex63}
\usepackage[utf8]{inputenc}

\usepackage{amsmath}
\usepackage{amssymb}
\usepackage{amsthm}

\usepackage{graphicx}

\usepackage{natbib}

\usepackage{soul}
\usepackage{verbatim}
\definecolor{darkgreen}{rgb}{0.05,0.3,0.05}

\usepackage{lineno}
\begin{document}

\vspace*{-\headsep}\vspace*{\headheight}
{\footnotesize \hfill FERMILAB-PUB-25-0649-PPD}\\
\vspace*{-\headsep}\vspace*{\headheight}
{\footnotesize \hfill DES-2025-0905}

\title{Robust Measurement of Stellar Streams Around the Milky Way: Correcting Spatially Variable Observational Selection Effects in Optical Imaging Surveys}

\correspondingauthor{Kyle Boone (kboone@g.harvard.edu) and Peter Ferguson (pferguso@uw.edu)}
%\email{kkboone@wisc.edu}
%\email{kbechtol@wisc.edu}

\author[0009-0005-7207-3342]{K.K.~Boone}
\affil{Department of Physics, Harvard University, Cambridge, MA 02138, USA}
\affil{Physics Department, University of Wisconsin-Madison, 1150 University Avenue Madison, WI  53706, USA}

\author[0000-0001-6957-1627]{P.S.~Ferguson}
\affil{DiRAC Institute and the Department of Astronomy, University of Washington, Seattle, WA, USA}
\affil{Physics Department, University of Wisconsin-Madison, 1150 University Avenue Madison, WI  53706, USA}

\author[0000-0002-0690-1737]{M.~Tabbutt}
\affil{Physics Department, University of Wisconsin-Madison, 1150 University Avenue Madison, WI  53706, USA}

\author[0000-0001-8156-0429]{K.~Bechtol}
\affil{Physics Department, University of Wisconsin-Madison, 1150 University Avenue Madison, WI  53706, USA}
\affil{Legacy Survey of Space and Time, 933 North Cherry Avenue, Tucson, AZ 85721, USA}

\author[0000-0001-8670-4495]{T.-Y.~Cheng}
\affiliation{Kapteyn Astronomical Institute, University of Groningen, Landleven 12 (Kapteynborg, 5419), 9747 AD Groningen, The Netherlands}

\author[0000-0001-8251-933X]{A.~Drlica-Wagner}
\affiliation{Fermi National Accelerator Laboratory, P.O.\ Box 500, Batavia, IL 60510, USA}
\affiliation{Kavli Institute for Cosmological Physics, University of Chicago, Chicago, IL 60637, USA}
\affiliation{Department of Astronomy and Astrophysics, University of Chicago, Chicago, IL 60637, USA}

\author[0000-0002-9144-7726]{C.~E.~Mart{\'\i}nez-V{\'a}zquez}
\affiliation{Cerro Tololo Inter-American Observatory, NSF's National Optical-Infrared Astronomy Research Laboratory, Casilla 603, La Serena, Chile}

\author[0000-0001-9649-4815]{B.~Mutlu-Pakdil}
\affiliation{Department of Physics and Astronomy, Dartmouth College, Hanover, NH 03755, USA}

\author[0000-0003-1587-3931]{T.~M.~C.~Abbott}
\affiliation{Cerro Tololo Inter-American Observatory, NSF's National Optical-Infrared Astronomy Research Laboratory, Casilla 603, La Serena, Chile}
\author[0000-0001-5679-6747]{M.~Aguena}
\affiliation{INAF-Osservatorio Astronomico di Trieste, via G. B. Tiepolo 11, I-34143 Trieste, Italy}
\affiliation{Laborat\'orio Interinstitucional de e-Astronomia - LIneA, Av. Pastor Martin Luther King Jr, 126 Del Castilho, Nova Am\'erica Offices, Torre 3000/sala 817 CEP: 20765-000, Brazil}
\author[0000-0002-7394-9466]{O.~Alves}
\affiliation{Department of Physics, University of Michigan, Ann Arbor, MI 48109, USA}
\author[0000-0003-0171-6900]{F.~Andrade-Oliveira}
\affiliation{Physik-Institut, University of Zürich, Winterthurerstrasse 190, CH-8057 Zürich, Switzerland}
\author[0000-0002-2562-8537]{D.~Bacon}
\affiliation{Institute of Cosmology and Gravitation, University of Portsmouth, Portsmouth, PO1 3FX, UK}
\author[0000-0002-4900-805X]{S.~Bocquet}
\affiliation{University Observatory, LMU Faculty of Physics, Scheinerstr. 1, 81679 Munich, Germany}
\author[0000-0002-8458-5047]{D.~Brooks}
\affiliation{Department of Physics \& Astronomy, University College London, Gower Street, London, WC1E 6BT, UK}
\author[0000-0002-7436-3950]{R.~Camilleri}
\affiliation{School of Mathematics and Physics, University of Queensland,  Brisbane, QLD 4072, Australia}
\author[0000-0003-3044-5150]{A.~Carnero~Rosell}
\affiliation{Instituto de Astrofisica de Canarias, E-38205 La Laguna, Tenerife, Spain}
\affiliation{Laborat\'orio Interinstitucional de e-Astronomia - LIneA, Av. Pastor Martin Luther King Jr, 126 Del Castilho, Nova Am\'erica Offices, Torre 3000/sala 817 CEP: 20765-000, Brazil}
\affiliation{Universidad de La Laguna, Dpto. Astrofísica, E-38206 La Laguna, Tenerife, Spain}
\author[0000-0002-7731-277X]{L.~N.~da Costa}
\affiliation{Laborat\'orio Interinstitucional de e-Astronomia - LIneA, Av. Pastor Martin Luther King Jr, 126 Del Castilho, Nova Am\'erica Offices, Torre 3000/sala 817 CEP: 20765-000, Brazil}
\author[0000-0002-7131-7684]{M.~E.~da Silva Pereira}
\affiliation{Hamburger Sternwarte, Universit\"{a}t Hamburg, Gojenbergsweg 112, 21029 Hamburg, Germany}
\author[0000-0002-4213-8783]{T.~M.~Davis}
\affiliation{School of Mathematics and Physics, University of Queensland,  Brisbane, QLD 4072, Australia}
\author[0000-0001-8318-6813]{J.~De~Vicente}
\affiliation{Centro de Investigaciones Energ\'eticas, Medioambientales y Tecnol\'ogicas (CIEMAT), Madrid, Spain}
\author[0000-0002-0466-3288]{S.~Desai}
\affiliation{Department of Physics, IIT Hyderabad, Kandi, Telangana 502285, India}
\author[0000-0002-6397-4457]{P.~Doel}
\affiliation{Department of Physics \& Astronomy, University College London, Gower Street, London, WC1E 6BT, UK}
\author[0000-0002-3745-2882]{S.~Everett}
\affiliation{California Institute of Technology, 1200 East California Blvd, MC 249-17, Pasadena, CA 91125, USA}
\author[0000-0002-2367-5049]{B.~Flaugher}
\affiliation{Fermi National Accelerator Laboratory, P. O. Box 500, Batavia, IL 60510, USA}
\author[0000-0003-4079-3263]{J.~Frieman}
\affiliation{Department of Astronomy and Astrophysics, University of Chicago, Chicago, IL 60637, USA}
\affiliation{Fermi National Accelerator Laboratory, P. O. Box 500, Batavia, IL 60510, USA}
\affiliation{Kavli Institute for Cosmological Physics, University of Chicago, Chicago, IL 60637, USA}
\author[0000-0002-9370-8360]{J.~Garc\'ia-Bellido}
\affiliation{Instituto de Fisica Teorica UAM/CSIC, Universidad Autonoma de Madrid, 28049 Madrid, Spain}
\author[0000-0003-3270-7644]{D.~Gruen}
\affiliation{University Observatory, LMU Faculty of Physics, Scheinerstr. 1, 81679 Munich, Germany}
\author[0000-0003-0825-0517]{G.~Gutierrez}
\affiliation{Fermi National Accelerator Laboratory, P. O. Box 500, Batavia, IL 60510, USA}
\author[0000-0003-2071-9349]{S.~R.~Hinton}
\affiliation{School of Mathematics and Physics, University of Queensland,  Brisbane, QLD 4072, Australia}
\author[0000-0002-9369-4157]{D.~L.~Hollowood}
\affiliation{Santa Cruz Institute for Particle Physics, Santa Cruz, CA 95064, USA}
\author[0000-0002-6550-2023]{K.~Honscheid}
\affiliation{Center for Cosmology and Astro-Particle Physics, The Ohio State University, Columbus, OH 43210, USA}
\affiliation{Department of Physics, The Ohio State University, Columbus, OH 43210, USA}
\author[0000-0001-5160-4486]{D.~J.~James}
\affiliation{Center for Astrophysics $\vert$ Harvard \& Smithsonian, 60 Garden Street, Cambridge, MA 02138, USA}
\author[0000-0003-0120-0808]{K.~Kuehn}
\affiliation{Australian Astronomical Optics, Macquarie University, North Ryde, NSW 2113, Australia}
\affiliation{Lowell Observatory, 1400 Mars Hill Rd, Flagstaff, AZ 86001, USA}
\author[0000-0003-0710-9474]{J.~L.~Marshall}
\affiliation{George P. and Cynthia Woods Mitchell Institute for Fundamental Physics and Astronomy, and Department of Physics and Astronomy, Texas A\&M University, College Station, TX 77843,  USA}
\author[0000-0001-9497-7266]{J. Mena-Fern{\'a}ndez}
\affiliation{Universit\'e Grenoble Alpes, CNRS, LPSC-IN2P3, 38000 Grenoble, France}
\author[0000-0002-1372-2534]{F.~Menanteau}
\affiliation{Center for Astrophysical Surveys, National Center for Supercomputing Applications, 1205 West Clark St., Urbana, IL 61801, USA}
\affiliation{Department of Astronomy, University of Illinois at Urbana-Champaign, 1002 W. Green Street, Urbana, IL 61801, USA}
\author[0000-0002-6610-4836]{R.~Miquel}
\affiliation{Instituci\'o Catalana de Recerca i Estudis Avan\c{c}ats, E-08010 Barcelona, Spain}
\affiliation{Institut de F\'{\i}sica d'Altes Energies (IFAE), The Barcelona Institute of Science and Technology, Campus UAB, 08193 Bellaterra (Barcelona) Spain}
\author[0000-0001-6145-5859]{J.~Myles}
\affiliation{Department of Astrophysical Sciences, Princeton University, Peyton Hall, Princeton, NJ 08544, USA}
\author[0000-0003-2120-1154]{R.~L.~C.~Ogando}
\affiliation{Observat\'orio Nacional, Rua Gal. Jos\'e Cristino 77, Rio de Janeiro, RJ - 20921-400, Brazil}
\author[0000-0002-2598-0514]{A.~A.~Plazas~Malag\'on}
\affiliation{Kavli Institute for Particle Astrophysics \& Cosmology, P. O. Box 2450, Stanford University, Stanford, CA 94305, USA}
\affiliation{SLAC National Accelerator Laboratory, Menlo Park, CA 94025, USA}
\author[0000-0002-2762-2024]{A.~Porredon}
\affiliation{Centro de Investigaciones Energ\'eticas, Medioambientales y Tecnol\'ogicas (CIEMAT), Madrid, Spain}
\affiliation{Ruhr University Bochum, Faculty of Physics and Astronomy, Astronomical Institute, German Centre for Cosmological Lensing, 44780 Bochum, Germany}
\author[0000-0001-6163-1058]{M.~Rodr\'{i}guez-Monroy}
\affiliation{Instituto de F\'{i}sica Te\'{o}rica UAM/CSIC, Universidad Aut\'{o}noma de Madrid, 28049 Madrid, Spain}
\affiliation{Laboratoire de physique des 2 infinis Ir\`ene Joliot-Curie, CNRS Universit\'e Paris-Saclay, Bât. 100, F-91405 Orsay Cedex, France}
\author[0000-0002-9646-8198]{E.~Sanchez}
\affiliation{Centro de Investigaciones Energ\'eticas, Medioambientales y Tecnol\'ogicas (CIEMAT), Madrid, Spain}
\author[0000-0003-3054-7907]{D.~Sanchez Cid}
\affiliation{Centro de Investigaciones Energ\'eticas, Medioambientales y Tecnol\'ogicas (CIEMAT), Madrid, Spain}
\affiliation{Physik-Institut, University of Zürich, Winterthurerstrasse 190, CH-8057 Zürich, Switzerland}
\author[0000-0002-1831-1953]{I.~Sevilla-Noarbe}
\affiliation{Centro de Investigaciones Energ\'eticas, Medioambientales y Tecnol\'ogicas (CIEMAT), Madrid, Spain}
\author[0000-0002-3321-1432]{M.~Smith}
\affiliation{Physics Department, Lancaster University, Lancaster, LA1 4YB, UK}
\author[0000-0002-7047-9358]{E.~Suchyta}
\affiliation{Computer Science and Mathematics Division, Oak Ridge National Laboratory, Oak Ridge, TN 37831}
\author[0000-0002-1488-8552]{M.~E.~C.~Swanson}
\affiliation{Center for Astrophysical Surveys, National Center for Supercomputing Applications, 1205 West Clark St., Urbana, IL 61801, USA}
\author{V.~Vikram}
\affiliation{Department of Physics, Central University of Kerala, 93VR+RWF, CUK Rd, Kerala 671316, India}
\author[0000-0001-9382-5199]{N.~Weaverdyck}
\affiliation{Berkeley Center for Cosmological Physics, Department of Physics, University of California, Berkeley, CA 94720, US}
\affiliation{Lawrence Berkeley National Laboratory, 1 Cyclotron Road, Berkeley, CA 94720, USA}

\collaboration{100}{(DES Collaboration)}

\begin{abstract}

Observations of density variations in stellar streams are a promising probe of low-mass dark matter substructure in the Milky Way. 
However, survey systematics such as variations in seeing and sky brightness can also induce artificial fluctuations in the observed densities of known stellar streams. 
These variations arise because survey conditions affect both object detection and star--galaxy misclassification rates.
To mitigate these effects, we use \texttt{Balrog} synthetic source injections in the Dark Energy Survey (DES) Y3 data to calculate detection rate variations and classification rates as functions of survey properties.
We show that these rates are nearly separable with respect to survey properties and can be estimated with sufficient statistics from the synthetic catalogs.
Applying these corrections reduces the standard deviation of relative detection rates across the DES footprint by a factor of five, and our corrections significantly change the inferred linear density of the Phoenix stream when including faint objects. 
Additionally, for artificial streams with DES like survey properties we are able to recover density power spectra with reduced bias.
We also find that uncorrected power-spectrum results for LSST-like data can be around five times more biased, highlighting the need for such corrections in future ground based surveys.

\end{abstract}

\keywords{Stellar streams, Cosmology, Sky surveys, Milky Way dark matter halo, Dark Matter}

\section{Introduction}
\label{sec:introduction}
The fundamental nature of dark matter is an outstanding question in physics and astronomy.
The standard model of dark energy plus cold dark matter ($\Lambda$CDM) fits data at large scales   and predicts the existence of smaller dark matter sub-halos in our Galaxy \citep[e.g.,][]{Bullock:2017, Buckley:2018, Chabanier:2019,  Bechtol:2022}. 
At these smaller scales the population of Milky Way satellites provides insight into the nature of dark matter down to the threshold of star formation ($M\sim 10^8M_\odot$), helping to constrain the microphysics of dark matter models and abundance of low mass subhalos.
Currently, these observations are consistent with the standard model \citep[e.g.,][]{Jethwa:2018, Nadler:2020, Newton:2021, Dekker:2022}. 
To continue to stress test $\Lambda$CDM and detect the impact of fully dark subhalos, we must look to additional probes. 
In the far-field, galaxy-scale gravitational lenses are expected to be sensitive to these dark subhalos down to a mass of $\sim 10^7 M_\odot$; either through flux-ratio anomalies, image positions, or time delays \citep[][and references therein]{Treu:2010, Vegetti:2024}.
Additionally, a promising near-field probe of dark subhalos comes from stellar streams around the Milky Way.

Stellar streams are the tidally disrupting remnants of star clusters and satellite galaxies \citep{Newberg:2016}. 
The discovery and characterization of these systems around the Milky Way has been enabled by wide-field astronomical surveys, initially through matched-filter searches of photometric data (e.g., SDSS; \citealt{Odenkirchen:2001}, Pan-STARRs; \citealt{Bernard:2014}, \citealt{Grillmair:2017}, DES; \citealt{Shipp:2018}), and subsequently using combinations of photometric, astrometric and spectroscopic data (e.g., Gaia; \citealt{Malhan:2018}, S5; \citealt{Li:2019}).
Currently, there are more than 120 identified stellar streams around our Galaxy \citep{Mateu:2023, Bonaca:2025}.

Within an individual stream, the stellar positions and velocities are probes of the local acceleration field experienced by the stream stars as they orbit around the Milky Way's gravitational potential \citep{Bovy:2017}.  
Therefore, in a Milky Way potential absent of substructure, it is expected that streams would show less substructure; although some will be present due to variations in the stripping rate and the formation of epicyclic overdensities \citep{Kupper:2010}. 
But, the presence of dark subhalos passing by a stream will induce perturbations in the on-sky positions, density, and velocity distribution of stream stars.  This can result in the formation of gaps and other small scale deviations in the stream track (e.g., \citealt{Erkal:2016}, \citealt{Bonaca:2019}, \citealt{Banik:2021a}, \citealt{Banik:2021b}, \citealt{Delos:2022}). 

The population of kinematically cold stellar streams (i.e., ones with globular cluster progenitors) comprise the best probe of this effect due to their small intrinsic velocity dispersions and stream widths \citep{Lu:2025}.
One such kinematically cold stellar stream is the Phoenix stream. 
This substructure, originally discovered in early DES data \citep{Phoenix_Discovery}, has a length of 4.6 kpc ($\sim15^{\circ}$), a width of $~0.14^{\circ}$, and a measured velocity dispersion of $\sigma_{rv}=2.66 \,\mathrm{km/s}$ \citep{Shipp:2018, Wan:2020}. 
Interestingly, this stream also shows small scale density fluctuations, making it a promising candidate for probing the low mass regime of the subhalo mass function \citep{Phoenix}.
Due to these characteristics, we use this stream as an example application in our analysis.
For thin streams in general, there has been a large effort to follow up discoveries using deeper imaging surveys to access fainter stars and provide the best constraints on the track and density variations \citep[e.g.,][]{Erkal:2017, Koposov:2019, Bonaca:2020, Li:2021, Phoenix, Patrick:2022}, while proper motion measurements \citep{Shipp:2019} and targeted spectroscopy of brighter members \citep{ Bonaca:2021,Li:2022} can be used to refine the stellar samples, constrain dynamics and further characterize the interaction history of these substructures. 

One common approach to quantify the observable effects of gravitational interactions with dark matter is through power-spectrum-type analyses of density variations (stars/deg) along a stream \citep{Banik:2021a, Delos:2022}.
Such inferences rely on accurate measurements of the density and track along the stream.
At small spatial scales, shot noise from the limited number of stars dominates the stream density power spectrum, thus it is most important to reduce bias on measurements at larger scales \citep[tens of deg;][]{Bovy:2017}. 

In photometric imaging data, systematic biases can be produced by unaccounted differences between intrinsic and observed source populations, i.e., the observational selection function.
The observational selection function depends on intrinsic source properties (e.g., flux in multiple photometric bands, surface brightness) together with external \emph{survey properties} (see Table \ref{tab:sp_maps}) that include both astrophysical effects (e.g., projected source density, interstellar extinction) and observational details (e.g., integrated exposure time) that generally vary over the survey footprint.
Streams covering tens of degrees on the sky will span multiple telescope fields of view, and are likely to be observed with a variety of survey properties. 

In addition, for most optical surveys, the observations are taken over multiple epochs under different observing conditions.
This produces variations in detection rates for both stars and galaxies in imaging data.
It also causes variations in how often stars and galaxies are incorrectly classified.
Already with DES Year 3 data, we find that spatially variable observational selection effects across the survey footprint induce statistically significant effects (Section \ref{sec:des_app}).
As our data and analyses become more sophisticated, percent-level variations in the survey-transfer function can induce systematic errors that are much larger than the statistical errors and therefore limit the power of analyses \citep{Balrog_Y3}. 

 Traditionally, to avoid these systematic biases, a selection for high signal-to-noise objects is applied to minimize variations in observational selection effects.
 For example, in the DES DR1 stream search, the stellar selection was limited to $g < 23.5$ \citep{Shipp:2018}.
 However, at the magnitudes that will be accessible to future surveys, such as the NSF-DOE Vera C. Rubin Observatory's Legacy Survey of Space and Time \citep[LSST, $5\sigma\, g$-band depth $\sim 27.4$ corresponding to around 2.3 magnitudes deeper than DES DR1;][]{LSST}, galaxies are much more common, and even small variations in galaxy misclassification rates could produce large effects on the stellar sample \citep{Gala_Impact}.
\emph{Therefore, new tools are needed to fully leverage the potential LSST-like data.}

In the context of large-scale structure analyses for cosmology, galaxy weight maps have been derived to account for spatially variable observational selection effects that would otherwise produce systematic errors in galaxy clustering measurements. % (e.g., 2-point correlation function).
These weight maps are derived using the ansatz that galaxies are isotropically distributed on large angular scales, and that empirical relations between observed galaxy densities and survey properties (e.g., seeing, integrated exposure time) can be used to learn the observational selection function for galaxies across the survey footprint \citep[see Section 5.3 of ][]{Rodriguez-Monroy:2022}. 
This approach is not possible for stellar samples, because the intrinsic distribution of stars across the survey footprint is highly non-isotropic.

In this work, we explore the use of synthetic source injection (SSI) in combination with survey metadata (i.e., survey property maps) to correct stellar samples for spatially variable observational selection effects. % Kyle:I think this is already implied. that would impact dark matter inference from stellar streams.
The SSI pipeline inserts realistic artificial stars and galaxies directly into pixel-level image data and re-runs source detection and measurement algorithms to effectively sample the observational selection function at locations across the survey footprint.
Specifically, we use the \texttt{Balrog} implementation of SSI for DES Y3 as a testing ground to develop the methodology \citep{Balrog_Y3}.% (Section 2).

The paper is organized as follows: in Section \ref{sec:data} we discuss the data from DES Y3 and \texttt{Balrog} that will be used in our corrections.
In Section \ref{sec:methods} we discuss the calculations involved in our corrections.
In Section \ref{sec:des_app} we apply our corrections to the full DES Y3 footprint and compare these results to pre-corrected data.
In Section \ref{sec:validation} we test the overall corrective power of our algorithm, how this power changes with larger training sets, and how repeatable our algorithm is.
In Section \ref{sec:str_app} we apply our corrections to observations of the Phoenix stream and investigate the resulting changes in linear density.
We then apply our corrections to simulated stellar streams and compare density power spectra.
Finally, we conclude and motivate potential future works in Section \ref{sec:conclusion}.
All code used in this project is available on github\footnote{\href{https://github.com/Kyle-Boone/ssi_corrections_des_y3_balrog}{\texttt{https://github.com/Kyle-Boone/ssi\_corrections\_des\_y3\_balrog}}} and Zenodo \citep{kyle_boone_2026_18662142}.

\section{Data}
\label{sec:data}
In this section we describe the observations, DES Y3 Gold catalog (Section \ref{sec:data_des}), and synthetic source injection runs, Balrog (Section \ref{sec:data_balrog}), used in our analysis.
\subsection{DES DR1 \& Y3 GOLD}
\label{sec:data_des}
\begin{table}
    \centering
    \begin{tabular*}{\columnwidth}{@{\extracolsep{\fill}}lll@{}}
        \hline
        \textbf{Quantity} & \textbf{Units} & \textbf{Statistics} \\ 
        \hline
        \textit{airmass} & --- & WMEAN \\ 
         &  & MIN \\ 
         &  & MAX \\ \hline 
        \textit{fwhm} & arcsec & WMEAN \\ 
         &  & MIN \\ 
         &  & MAX \\ \hline 
        \textit{fwhm\_fluxrad} & arcsec & WMEAN \\ 
         &  & MIN \\ 
         &  & MAX \\ \hline
        \textit{exptime} & seconds & SUM \\ \hline 
        \textit{t\_eff} & --- & WMEAN \\ 
         &  & MIN \\ 
         &  & MAX \\ \hline
        \textit{t\_eff\_exptime} & seconds & SUM \\ \hline
        \textit{skybrite} & e$^-$/CCD pix & WMEAN \\ \hline
        \textit{skyvar} & (e$^-$/CCD pix)$^2$ & WMEAN \\ 
         &  & MIN \\ 
         &  & MAX \\ \hline
        \textit{skyvar\_sqrt} & e$^-$/CCD pix & WMEAN \\ \hline
        \textit{skyvar\_uncertainty} & e$^-$/s$\cdot$coadd pix &  \\ \hline
        \textit{sigma\_mag\_zero} & mag & QSUM \\ \hline
        \textit{fgcm\_gry} & mag & WMEAN \\ 
         &  & MIN \\ \hline
        \textit{stellar\_dens} & stars/$\deg^2$ & ---  \\ 
        \hline
    \end{tabular*}
    \caption{
    These are the survey properties used in the analysis, along with their units and the different statistics used. Each property has maps in $griz$ bands with the exception of \textit{stellar\_dens} and \textit{skyvar\_sqrt} (which was lacking an $r$-band map). Adding up all the statistics and bands across each each quantity gives a total of 92 maps used.}
    \label{tab:sp_maps}
\end{table}

\begin{figure*}
    \centering
     \includegraphics[width=0.90\textwidth]{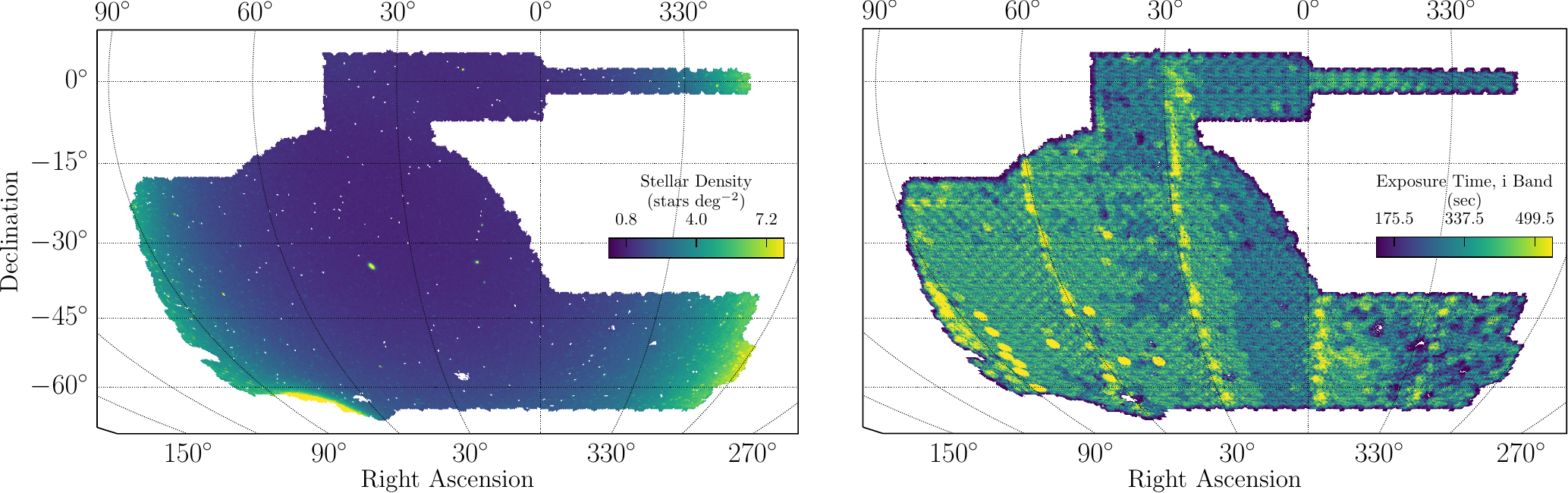}
      \caption{Distributions of two of the survey properties used, stellar density (left) and exposure time sum in the $i$-band (right).} 
      \label{fig:two_props}
\end{figure*}

%This analysis makes use of 
%We use data products from the first data release of DES \citep[DR1;][]{DES_DR1}.
%This data was used instead of more recent catalogs since the synthetic sources we use were injected into this catalog.
We use data products from the first data release of DES \citep[DR1;][]{DES_DR1} based on three years of observations using the Dark Energy Camera \citep{DeCam} mounted on the Blanco 4m telescope at the Cerro Tololo Inter-American Observatory (CTIO).
%in Chile
%DES DR1 includes three years of observations using the Dark Energy Camera \citep{DeCam} mounted on the Blanco 4m telescope in Chile at the Cerro Tololo Inter-American Observatory (CTIO).
The survey data covers an area of $\sim 5000\,\,\text{deg}^2$ in five broadband filters, $grizY$. 
With 38,850 total exposures in DES DR1, each location in the footprint typically contains $4-6$ exposures in each band \citep{Diehl:2016}, corresponding to a median depth of $i\sim 23.3$ at $S/N = 10$ for unresolved sources.
%In the $i$-band corresponds to a median depth of $i\sim 23.3$ for S/N $= 10$ point sources.

% The Dark Energy Survey collected imaging data with the Dark Energy Camera \citet{DeCam} over six years: from 2013 to 2019.
% The camera was mounted on the Blanco 4m telescope in Chile at the Cerro Tololo Inter-American Observatory (CTIO).
% During the survey, an area of $\sim 5000\,\,\text{deg}^2$ was observed on the sky in five broadband filters, grizY.
% The analysis in this work used data from the first data release \citet{DES_DR1} which accounted for the first three years of observing. This had an effective depth of $i\sim 23.3$ for S/N $= 10$ point sources.

We also use object classifications, quality flags, and survey property maps from the value-added DES Y3 GOLD release \citep{Y3_Gold}. 
%Additionally, we make use of the Y3 GOLD \citep{Y3_Gold} value added products. 
%In particular, we use survey property maps and quality flags.
The survey property maps consist of spatial survey property values stored as HEALPix\footnote{\href{http://healpix.sourceforge.net}{http://healpix.sourceforge.net}} \citep{HEALPix} pixel (hereafter healpixel) maps.
These survey property maps track spatial variations of observing conditions at a high resolution (\texttt{NSIDE} = 4096) and are described in more detail in Section 7.3 and Appendix E of \citet{Y3_Gold}. 
Table \ref{tab:sp_maps} lists the survey properties that are used in this analysis.
Examples of survey property maps for stellar density and exposure time in the $i$-band are shown in Figure \ref{fig:two_props}.
%Since our survey properties are stored on HEALPix pixels, these are the spatial pixels we perform corrections on.

For the flags, to obtain a high quality sample of objects we base our selection on  \citet{Balrog_Y3} (see their Section 4) and place the following cuts on all Y3 GOLD objects:

% Y3 GOLD \citet{Y3_Gold} adds additional information to the DR1 catalog.
% These additions include survey property maps and quality flags, both of which are used in this analysis.
% The survey properties used in this work are summarized in \ref{tab:sp_maps}.
% Throughout this work, the following cuts are made for all Y3 GOLD objects.

\begin{align}
  &\texttt{FLAGS\_FOREGROUND} = 0\nonumber\\
  \text{AND}\;\; & \texttt{FLAGS\_BADREGIONS} < 2 \nonumber\\
    \text{AND}\;\; & \texttt{FLAGS\_FOOTPRINT} = 1 \nonumber.
\end{align}

For the above cuts, the footprint cut is used to select regions which had good coverage in multiple observing bands, the foreground cut is used to remove regions near bright objects, and the bad regions cut is used to remove high densities of anomolous colors and tiles where the multi-object fitting pipeline failed to finish.
More details on the above cuts can be found in Section 7.1 and 7.2 of \citet{Y3_Gold}.

% These cuts follow section 4 of \citet{Balrog_Y3}.
Object classification is done using single object fitting (SOF) classification with $\texttt{EXTENDED\_CLASS\_SOF}$, which is the only classification provided within our synthetic source catalog.
Stars are defined as objects with $0 \leq \texttt{EXTENDED\_CLASS\_SOF} \leq 1$, galaxies are objects with $2 \leq \texttt{EXTENDED\_CLASS\_SOF} \leq 3$.

%\PSF{do we need to justify using SOF values and describe what those are?}
% Distinctions between stars and galaxies are made using $\text{EXTENDED\_CLASS\_SOF}$.
% In this work, we define stars as objects with $0 \leq \text{EXTENDED\_CLASS\_SOF} \leq 1$.
% Galaxies are defined as objects with $2 \leq \text{EXTENDED\_CLASS\_SOF} \leq 3$.

Object magnitudes are also incorporated in our methodology, in all cases we use the \texttt{SOF} magnitudes which are single-object multi-epoch measurements described in \citet{Y3_Gold}.
However, in this work point spread function ($\texttt{SOF\_PSF\_MAG}$) magnitudes are used for objects classified as stars, while composite model ($\texttt{SOF\_CM\_MAG}$) magnitudes \citep{CModel} are used for objects classified as galaxies.
%$\texttt{\_\{BAND\}}$ magnitudes are generated by finding the best fit of a linear combination of exponential and de Vaucouleurs fits \citep{deVaucouleurs}.

\subsection{Balrog Synthetic Sources}
\label{sec:data_balrog}
To compliment the Y3 GOLD data, the synthetic source catalog generated by \texttt{Balrog} \citep{Balrog_Y3, Original_Balrog} is used.
\texttt{Balrog} is a software package that synthetically injects sources into individual DES images before coaddition and processes them with the same DESDM pipelines as observations.
\texttt{Balrog} is the name used to refer to both the software and synthetic object catalog, but for the remainder of this work we will use this name only to refer to the catalog. 
The injected sources were chosen to be empirical populations of artificial galaxies and stars taken from the deeper DES Deep Field observations \citep{Deep_Fields}. 
Additionally, delta-function stars (delta functions convolved with a local PSF; Section 2.2.4 of \citet{Balrog_Y3}) are injected, which we use for our synthetic stellar sample to avoid any potential galaxy contamination from incorrect Deep Field classifications.
These synthetic objects are subject to the same systematic impacts from survey properties as physical objects would be.
For this work we use all \texttt{Balrog} objects that pass our quality cuts.
This consists of 7.4M total objects (90\% Deep Field objects, 10\% delta-stars).
These objects were injected on a uniform grid on 2,041 randomly chosen tiles out of the 10,338 total Y3 tiles (for a coverage map see Figure 6 of \citealt{Balrog_Y3}, and more information on the injected objects can be found in their Section 3).

In this work, \texttt{Balrog} objects are subject to the same flag cuts as described previously for Y3 GOLD objects.
Classification cutoffs in terms of $\texttt{EXTENDED\_CLASS\_SOF}$ are also the same.
Likewise, $\texttt{PSF}$ magnitudes are used for objects classified as stars while $\texttt{CM}$ magnitudes are used for objects classified as galaxies.
To be consistent with Y3 GOLD objects, measured magnitudes are used for objects instead of true magnitudes.
Finally, a cut of $\texttt{MATCH\_FLAG\_1.5\_ASEC} < 2\nonumber$ is applied, which reduces the risk of ambiguous matching to Y3 GOLD objects within $1.5$ arcseconds.
More details on this flag cut can be found in Section 3.5 of \citet{Balrog_Y3}. 
With this synthetic source injector, we are able to perform corrections on DES data, which we turn to next.

% one additional cut is made for Balrog objects, inspired once again by section 4 of \citet{Balrog_Y3}:

% \begin{equation}
%     \text{MATCH\_FLAG\_1.5\_ASEC} < 2\nonumber
% \end{equation}

\section{Methods}
\label{sec:methods}
\subsection{Overview and Notation}\label{overview}

\begin{table}
    \centering
    
    \begin{tabular*}{\columnwidth}{@{\extracolsep{\fill}}lll@{}}
        \textbf{Notation} & \textbf{Terminology} & \textbf{Definition} \\ \hline

        $\text{\_}_S$ & & Stars. \\ \hline

        $\text{\_}_G$ & & Galaxies. \\ \hline

        I\_ & & Injected objects. \\ \hline
        
        $T_S$ & True Stars & True number of stars \\
        & & in a given area. \\ \hline

        $O_S$ & Observed Stars & Count of detected true \\
        & & stars in a given area, \\
        & & regardless of given \\
        & & classification. \\ \hline

        $C_S$ & Classified Stars & Count of objects \\
        & & classified as stars \\
        & & in a given area. \\ \hline

        $P\left( C_S|O_S\right)$ & & Probability of giving \\
        & & an observed star a \\
        & & classification of star. \\ \hline

        $RO_S$ & Recovered & Recovered count for \\
        & Observed Stars & observed stars after \\
        & & applying maximum \\
        & & likelihood separation. \\ \hline

        $D_R\left( C_S, O_S\right)$ & & Relative detection rate \\
        & & of observed stars \\
        & & classified as stars. \\ \hline

        $FC_S$ & (Final) Corrected & Final star counts after \\
        & Stars & applying corrections. \\ \hline
    \end{tabular*}
    \caption{
    This is a description of notation and terminology that will be used throughout this paper. Lesser used notation will be defined when used.}
    \label{tab:notation}
\end{table}

Spatial variations in survey properties lead to correlated variations in the stellar selection function in two ways: (i) variations in the \emph{correct classification} rate of detected objects and (ii) variations in the \emph{object detection} rate. % as stars.
The notation used in this work is summarized in Table \ref{tab:notation}.
In particular, $T_S$ is the true number of stars, detected and undetected, in any area of the sky.
$O_S$ is the number of these true stars which are detected, and $C_S$ is the number of objects that are classified as stars (this includes misclassified galaxies).

Our algorithm uses SSI objects and survey property maps to derive a relation between the survey properties and the likelihood that a given object will be detected and classified correctly. 
This will allow us to take catalog level data ($C_S$ and $C_G$) and correct it based on the survey properties at a given on-sky location. 
Our correction will be based on the assumption that the synthetic objects will respond in the same way to survey property variations as real data.
For \texttt{Balrog} objects, this assumption is appropriate based on Figure 11 of \citet{Balrog_Y3}.
Since we know whether \texttt{Balrog} objects are truly stars or galaxies, we can use them to measure the rates at which objects are classified correctly.
This measurement is critical for getting high accuracy counts of stars at faint magnitudes since misclassified galaxies are the dominant contributor to uncorrected faint star counts.

To start our correction, we therefore first use SSI objects to obtain the probability that a detected SSI star/galaxy will be classified correctly ($P\left( IC_S|IO_S\right)$ and $P\left( IC_G|IO_G\right)$).
% Note that quantities with an $I$ in front hold their usual definition but with injected objects.
We assume that the probability that a detected star (galaxy) will be classified correctly is the same as that probability for an SSI star (galaxy):

\begin{equation}
\label{eq:prob_match}
    P\left( C_S|O_S\right) = P\left( IC_S|IO_S\right)
\end{equation}

These probabilities allow us to estimate the number of true stars (galaxies) that were detected $RO_S$ ($RO_G$) at each position. 
After obtaining $RO_S$ and $RO_G$, we use the synthetic sources to estimate \emph{relative} detection rates (for correctly classified stars $D_R\left( C_S, O_S\right)$ is the rate at which true stars $T_S$ enter the stellar sample $C_S$ relative to the average rate across the footprint).
We find that correctly classified stars are subject to \emph{distinct} variations in the relative detection rate compared to misclassified galaxies (see Appendix \ref{app:rdrc}), which necessitates the calculation of both for a full correction.
Therefore, we calculate four relative detection rates: $D_R\left( IC_S,IO_S\right)$, $D_R\left( IC_G,IO_S\right)$, $D_R\left( IC_S,IO_G\right)$, $D_R\left( IC_G,IO_G\right)$. 
%After obtaining $RO_S$ and $RO_G$, we use the synthetic sources to estimate the relative detection rates for four different groups of objects: SSI stars classified as stars $D_R\left( IC_S,IO_S\right)$, SSI stars classified as galaxies $D_R\left( IC_G,IO_S\right)$, SSI galaxies classified as stars $D_R\left( IC_S,IO_G\right)$, and SSI galaxies classified as galaxies $D_R\left( IC_G,IO_G\right)$. 
As in Eq.\ (\ref{eq:prob_match}), we assume the relative detection rates for real objects in these groups are the same as for SSI objects.

Classification probabilities and relative detection rates are used to obtain final corrected counts at each position ($FC_S$ \& $FC_G$). 
We present two different types of corrections: one estimates what $C_S$ would have been with uniform survey properties, and one estimates what $O_S$ would have been with uniform survey properties.
More details are given in Section \ref{Algo_Design}.

For large surveys, the stellar magnitude distribution is not uniform over the full footprint.
To account for this, we bin all sources based on magnitudes.
All probabilities, detection rates, and corrections are calculated for each magnitude bin.

\subsection{Probability Calculations}

\begin{figure}
  \includegraphics[width=0.45\textwidth]{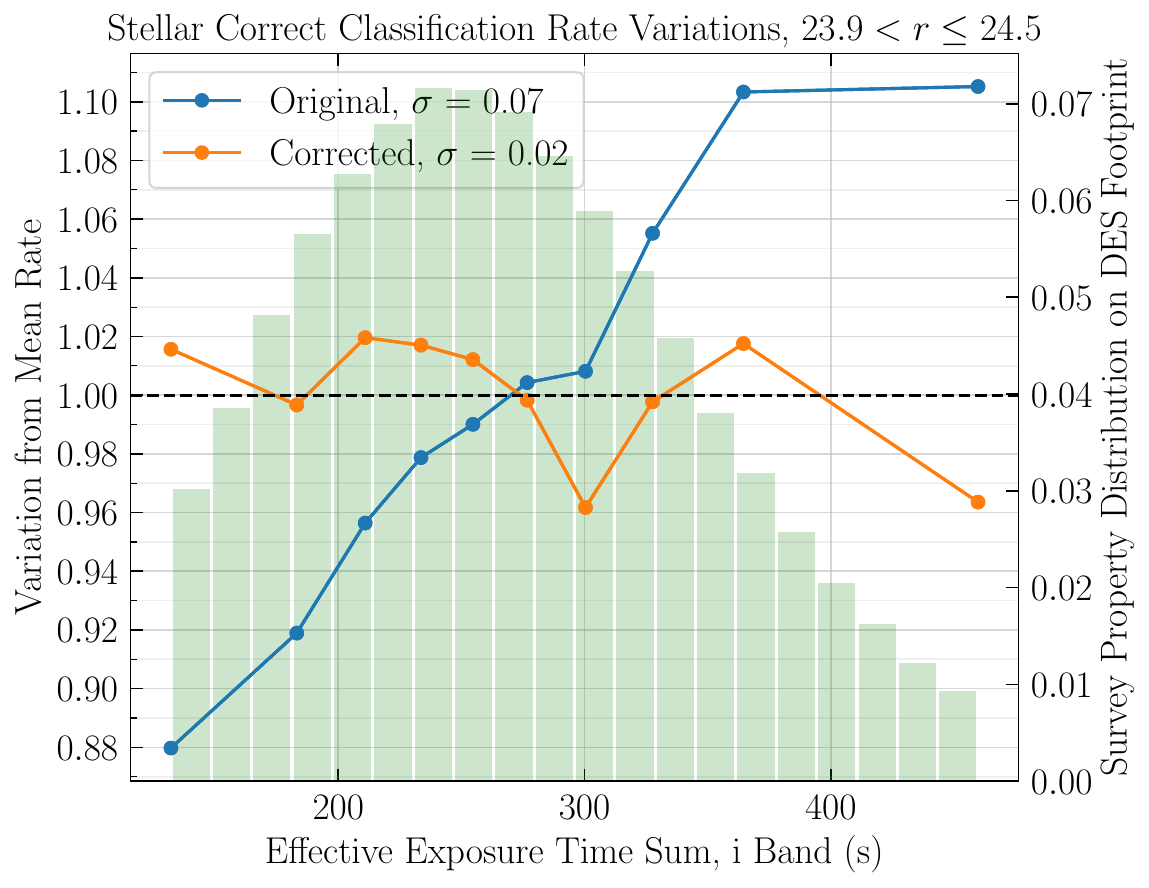}
  \caption{This figure shows \texttt{Balrog} delta star correct classification rates before (blue) and after (orange) corrections as a function of the effective exposure time sum in the $i$-band. The green histogram shows the distribution of effective exposure time in the $i$-band across the DES footprint. Training was performed on an 80\% subsample of the \texttt{Balrog} objects with the remaining sources used for testing. As well as magnitude cuts, a color cut to the Phoenix isochrone is made (see Section \ref{matched_filt}). A drop in standard deviation after corrections shows a mitigation of some of the variations caused by this survey property. \label{fig:classification_variations}}
\end{figure}

This section outlines the calculation of $P\left(C_S|O_S\right)$ and $D_R\left(C_S, O_S\right)$.
A more detailed description can be found in Appendices \ref{app:prob} and \ref{app:sep}.

As a specific example to illustrate our method, we present $P\left(IC_S|IO_S\right)$ in our faintest magnitude bin as a function of a single survey property in Figure \ref{fig:classification_variations}.
The green histogram shows the distribution of effective exposure time in the $i$-band for DES Y3. 
The blue (orange) line shows the ratio before (after) corrections of correctly classified SSI stars to all detected SSI stars as a function of the survey property value (Table \ref{tab:sp_maps}).
Since we only care about the variations in this rate, we divide each line by its average to center them at $1$.
For corrections, this ratio of correctly classified SSI stars to all detected SSI stars is computed for all survey properties.
We then take the survey property with the largest variation in the ratio and apply an empirical correction. 
This process is repeated iteratively until one of our termination criteria is met: either $\sigma < 0.01$ for all survey properties or, to avoid overfitting and/or long runtimes, after 150 iterations (see Section \ref{repeatability}).
To ensure that we are not just removing noise, we take 80\% of our SSI stars to be a training sample and the remaining 20\% to be a testing sample.
All corrections were calculated on the training sample but then applied to the testing sample to make the plot.
In this way, the drop in standard deviation from $\sigma=0.07$ to $\sigma = 0.02$ indicates that our correction pipeline has reduced the impact variations in this survey property have on the stellar classification rate.
We see $\sigma > 0.01$ after corrections since the test set is independent from the training set.

Relative detection rates are calculated in a similar way as classification probabilities, and details are left in Appendix \ref{app:prob}.
We find that variations in relative detection rates are larger than variations in classification probabilities.
Before corrections, Figure \ref{fig:classification_variations} shows that among all  bins $P\left(C_S|O_S\right)$ varies from the average by a maximum of $\sim 12$\%. 
For these same survey property bins, $D_R\left(C_S, O_S\right)$ varies from the average by a maximum of $\sim 39$\%, so we allow training to continue for 300 cycles before stopping if convergence had not been reached.

\subsection{Algorithm Design}\label{Algo_Design}

With classification probabilities and relative detection rates established, we now detail the correction algorithms introduced in Section \ref{overview}, which share identical steps except for their final stage.

The first step uses $P\left(C_S|O_S\right)$, $P\left(C_G|O_G\right)$, $C_S$, and $C_G$ to estimate the number of true stars (galaxies) that were detected, $RO_S$ ($RO_G$).
To start, we note that the expected number of classified stars is:

\begin{equation} 
    \langle C_S\rangle = O_SP\left(C_S|O_S\right) + O_G\left[ P\left(C_G|O_G\right) - 1\right],
\end{equation}
with an analogous equation for $\langle C_G\rangle$.
For the above equation, we note that $\left[ P\left(C_G|O_G\right) - 1\right]=P\left(C_S|O_G\right)$.

These equations can be inverted to solve for the expected number of observed stars and observed galaxies ($RO_S$ and $RO_G$).
This gives the maximum likelihood estimate for $RO_S$ and $RO_G$ of:

\begin{equation} \label{eq:ROS}
    RO_S = \frac{C_SP\left(C_G|O_G\right) + C_G\left[ P\left(C_G|O_G\right) - 1\right]}{P\left(C_S|O_S\right) + P\left(C_G|O_G\right) - 1}
\end{equation}

This value can be negative or larger than $C_S+C_G$, both of which are not physical. 
We crop $RO_S$ to be between zero and $C_S+C_G$.
Using conservation of counts, we get $RO_G$ by demanding that $RO_S+RO_G=C_S+C_G$.

All relative detection rates behave distinctly, so each one must be used to make corrections to specific subsets of the objects.
For more details, refer to Appendix \ref{app:rdrc}.
Due to this, our next step is getting estimates on the number of objects in the groups $C_S\cap RO_S$, $C_S\cap RO_G$, $C_G\cap RO_S$, and $C_G\cap RO_G$ (this is just the classification distribution for $RO_S$ and $RO_G$), given by:

\begin{equation}
\label{eq:CSROS}
    C_S\cap RO_S = RO_S P\left(C_S|O_S\right)
\end{equation}

Other combinations are calculated analogously.
Correcting any of these four sets of objects for variable detection rates is done by dividing by their respective relative detection rate, for example:% correcting $C_S\cap RO_S$ can be done by letting:

\begin{equation}
    \label{eq:corr_CSROS}
    \left( C_S\cap RO_S\right)_\text{Corr} = \frac{C_S\cap RO_S}{D_R\left( C_S,O_S\right)}
\end{equation}

Obtaining $FC_S$ (and $FC_G$) is now just a matter of algorithmic design.
We could take $FC_S$ to be the sum of the corrected counts for $C_S\cap RO_S$ and $C_S\cap RO_G$ as in Eq.\ (\ref{eq:FCS}).
This design tries to remove variations induced by survey properties in $C_S$.
Alternatively we could take $FC_S$ to be the sum of the corrected counts for $C_S\cap RO_S$ and $C_G\cap RO_S$ as in Eq.\ (\ref{eq:FCS2}).
This design tries to remove variations induced by survey properties in $O_S$.

\begin{equation}
\label{eq:FCS}
    FC_S = \left(C_S\cap RO_S\right)_\text{Corr} + \left(C_S\cap RO_G\right)_\text{Corr}
\end{equation}

\begin{equation}
\label{eq:FCS2}
    FC_S = \left(C_S\cap RO_S\right)_\text{Corr} + \left(C_G\cap RO_S\right)_\text{Corr}
\end{equation}

In this work, we use the first of these algorithms, Eq.\ (\ref{eq:FCS}), which is justified in Appendix \ref{app:adt}.

\section{Correction of the DES Data}
\label{sec:des_app}
In this section we present the results of applying our correction algorithm to DES data. 
To select a realistic dataset for this pipeline that will be applicable to a stream, we follow the study of the Phoenix stream by \citet{Phoenix}. % we follow \citet{Phoenix}'s study of the Phoenix stream. 
Phoenix is a $15^{\circ}$ long, $0.16^{\circ}$ wide, and dynamically cold stellar stream in the Southern Hemisphere \citep{Phoenix_Discovery, Phoenix_Endpoints, Phoenix}.
Phoenix shows density fluctuations on small scales, making it a well-suited candidate for studying potential perturbations.
In addition to its intrinsic properties, the stream lies in the middle of the DES footprint, near a prominent survey depth feature at a right ascension (RA) of $\sim 30$ deg (Figure \ref{fig:two_props}).
This feature has increased depth and is one of the most readily apparent survey property features in the entire DES footprint. 
Therefore, we can use observations of the Phoenix stream to demonstrate the application of our pipeline and its effect on the generation of density maps. 

Initially, we define a color-magnitude-based matched filter data selection used to obtain a realistic set of catalog objects.
Then, we discuss the adjustments needed to apply our method to this dataset and generate corrections for both stars and galaxies. 

\subsection{Matched Filter Data selection}
\label{matched_filt}

To derive corrections for a given stellar or galactic density map we want to place the same observational selection criteria on the injected objects as we would place on observations. 
In the case of the Phoenix stream, we use the matched filter from \citet{Phoenix}.
This filter is described in detail in their analysis, but generally it uses a synthetic isochrone to generate a selection region in color-magnitude space. 
The size of this region is defined by the expected width of the stellar population convolved with observational uncertainties. 
We use an isochrone from 
\citet{Bressan_Isochrone} as implemented in \texttt{ugali} \citep{Bechtol_Ugali,Drlica_Wagner_Ugali} with an age of a 
$=12.8$ Gyr, metallcity of [Fe/H] $=-2.5$, and distance of d $=17.4$ kpc \citep{Phoenix}.
In practice, extinction corrected data should be used to apply the isochrone cut and with apparent magnitudes then being used to apply corrections.
This could be achieved by adding an extinction map as an additional survey property, but more \texttt{Balrog} objects would be needed to cover the full range of apparent magnitude space necessary for corrections, and binning would likely have to be done by color as well as magnitude.
Since this paper is focused on proof of concept corrective power rather than legitimate DM constraints, we simplify our pipeline to only work with apparent magnitudes.
 
We also use a magnitude limit of $r < 24.5$, where completeness falls to $\sim20 \%$ (see \citealt{Balrog_Y3} Figure 9).
This can be compared to \citet{Shipp:2018} which use the same observations (DES Y3) but place a limit around one magnitude brighter ($g < 23.5$).
It is also $0.3$ mag fainter than the $r < 24.2$ limit used in \citet{Phoenix} which uses DES Y6 data (roughly twice as many observations as DES Y3). 
% This limit is one magnitude fainter than that of 

% allowed us to slightly deeper than the  signal to noise limit used in \citet{Phoenix} while staying at around a 20\% completeness, though we note that the \citet{Phoenix} analysis used roughly twice as many observations as ours 

As mentioned in Section 2.1, in our correction algorithm we apply corrections to different magnitude bins.
Our magnitude bins are as follows: $r \leq 22.9$, $22.9 < r \leq 23.9$, $23.9 < r \leq 24.5$.
These magnitude bins were chosen as they had similar relative detection rate variations in testing sets when these variations were extrapolated out to expected values for a more current \texttt{Balrog} run.
For more details, refer to Appendix \ref{app:mag}.
%Additionally we required magnitude limit of $g < 27$ is applied to ensure we would not get any objects that were too faint in this band.
These selections were placed in addition to the ones described in Section \ref{sec:data}.
% The heliocentric distance of Phoenix is nearly constant, with differences of only $0.13\pm 0.09$ kpc over the length of the stream \citep{Phoenix}.
% This makes for a simpler isochrone selection for Balrog objects than if the heliocentric distance was more varied.
% On top of the relative simplicity from the low distance gradient, Phoenix is located 
Throughout this section we show figures for these magnitude bins only, as this best illustrates the effectiveness of our method in extending the usable magnitude range for stream analyses; at brighter magnitudes the catalog is significantly more uniform.

% The goal of this pipeline is applying these corrections to stellar streams, so we want to use synthetic sources with similar properties to the stellar stream stars.
% This means selecting Balrog objects which fall along the isochrone of the target stellar stream.
% For this work, we use a synthetic isochrone \citep{Bressan_Isochrone} implemented in \texttt{ugali} \citep{Bechtol_Ugali,Drlica_Wagner_Ugali}.
% Parameters for our selected isochrone were chosen in accordance with \citet{Phoenix}.
% On top of this isochrone cut, an overall upper limit on the r band magnitude was placed at 24.5.
% This cut was 

\subsection{Algorithm Adjustments for Real Data}
%We make the following changes to apply our algorithm to observational data. 
Due to noise concerns, when applying to observational data, we perform the corrections at a healpixel resolution of \texttt{NSIDE} = 512. 
%The classification and detection probabilities are computed at a higher resolution of \texttt{NSIDE} = 4096, but the maximum likelihood estimation is more stable with larger object counts in each healpixel.
A \texttt{FRACDET} map \citep{Y3_Gold} is used for a weighted resolution degradation and for a first order correction to counts.
We crop to healpixels that have a \texttt{FRACDET} value $>$ 0.5.
%This map consists of the percentages of the \texttt{NSIDE} = 4096 healpixels that were observed in all $griz$ bands.
%To degrade the probability maps to a \texttt{NSIDE} = 512 resolution we use a weighted average of the \texttt{NSIDE} = 4096 resolution map, where the weights are given by a \texttt{FRACDET} map. 
%We also divide the \texttt{NSIDE} = 512 object count maps by these same \texttt{FRACDET} weights as a first correction.
%To ensure the healpixels we use have a reasonable coverage, we crop to healpixels that have a \texttt{FRACDET} value $>$ 0.5.

% While probabilities were calculated at a 4096 healpixel resolution, the maximum likelihood aspect of the pipeline works best when large object counts are present.
% Therefore, a FracDet map (the fraction of each 4096 resolution pixel which was detected) was used to degrade each probability map to a new resolution of 512.
% The FracDet map served as a weighting value for the subpixels of each new 512 resolution healpixel.
% We also immediately corrected star and galaxy counts for FracDet variations by dividing counts by the FracDet value on that pixel.

In this work, Deep Field object classifications are taken as the truth.
By performing spatial matching between the DES Deep Fields and Wide Fields and looking at the classifications given in both cases, we are able to get true correct classification rates.
When comparing these rates against the classification rates in that area based on our training we found systematic errors.
To correct this we multiplied our classification probabilities by scalars to match the probabilities shown by the deep fields.
%KYLE 2025: I don't think this information is necessary.
%In the brightest ($r \leq 22.9$), middle ($22.9 < r \leq 23.9$), and faintest ($23.9 < r \leq 24.5$) magnitude bins respectively our correct classification probabilities for stars were multiplied by $0.994$, $0.936$, and $0.875$.
%For those same magnitude bins our correct classification probabilities for galaxies were multiplied by $1.006$, $1.015$, and $1.037$.

%On top of this, slight adjustments are made to our classification probabilities.
%Slight mismatches in Balrog properties and real object properties can lead to the average classification probabilities being off.
%To account for this potential issue, we calibrate our classification probabilities off of the deep fields.
%For wide field objects that can be matched to deep field objects, we have a true classification from the deep fields and a classification from the DES pipeline.
%Using this we can get a correct classification rate in the deep fields, compare that to what our calculated classification rates are in that area, and use the difference to calibrate our calculated classification probabilities.

\subsection{DES Corrections}

\begin{figure*}
    \centering
     \includegraphics[width=0.90\textwidth]{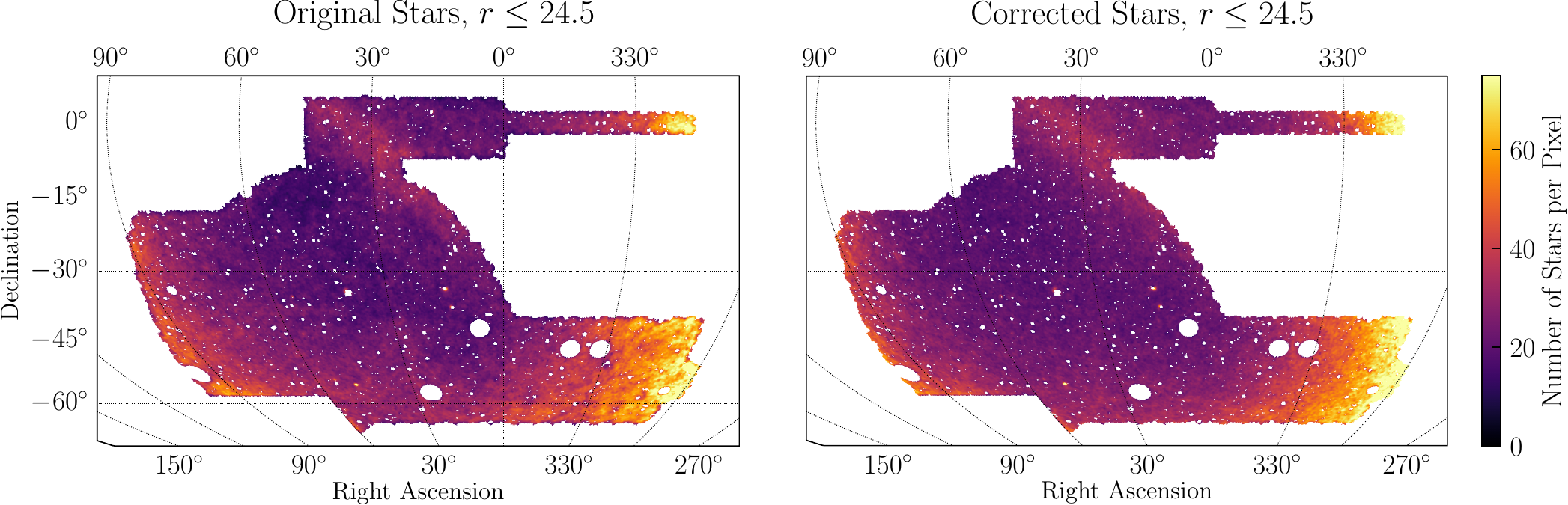}
      \caption{Star counts before and after corrections are applied. The matched filter described in Section 4.1 is used to crop for color and magnitude. Masked regions are primarily masked due to bright foreground objects, which is also true for future plots. For more physical units, at the above \texttt{NSIDE} = 512 each pixel has an area of $\sim 47.2 \text{ arcmin}^2$. Notable improvements are the suppression of the depth feature at RA $\sim 30^\circ$ and the better continuity at the edge of the footprint, both of which are more obvious in the galaxy plot below.} 
      \label{fig:Star_Correction}
\end{figure*}

\begin{figure*}
    \centering
     \includegraphics[width=0.90\textwidth]{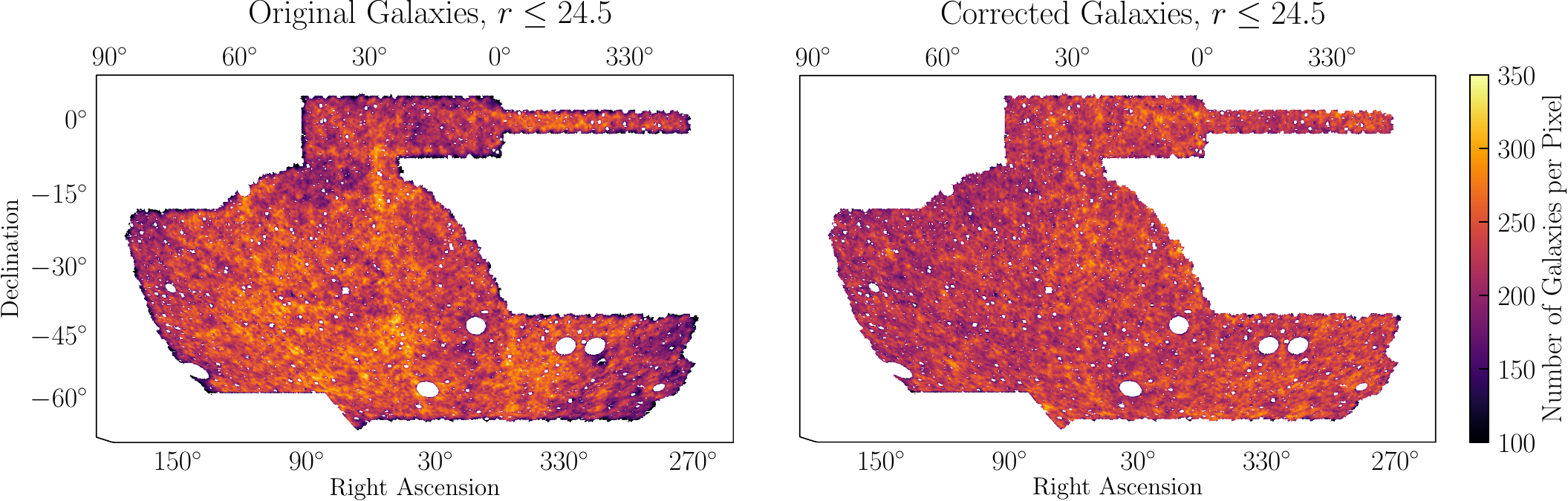}
      \caption{Galaxy counts before and after corrections are applied.} 
      \label{fig:Galaxy_Correction}
\end{figure*}

\begin{figure}
  \includegraphics[width=0.45\textwidth]{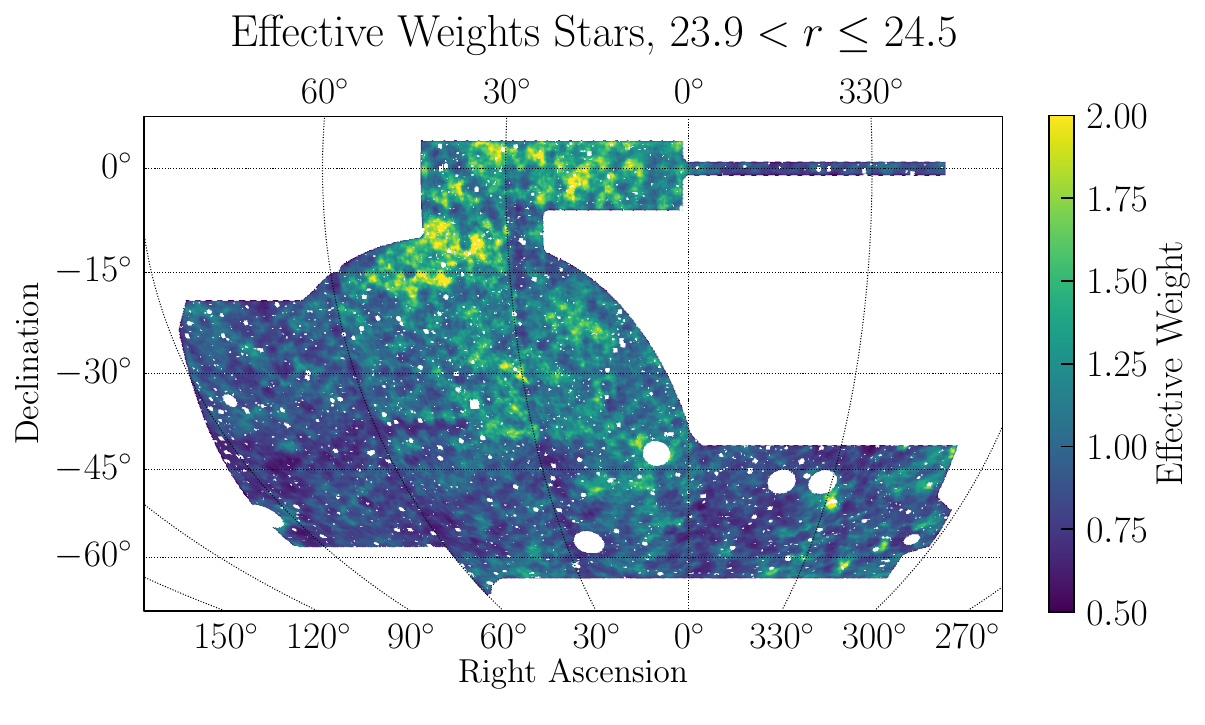}
  \caption{The effective weight map (corrected / uncorrected), smoothed with a $0.15^{\circ}$ kernel, for stellar objects in our faintest magnitude bin. This map is used for the validation tests in Section \ref{sec:validation}.}
  \label{fig:weights}
\end{figure}

% Galaxy counts can provide a probe as to how well the algorithm is working since they should be uniform on large scales across the DES footprint.
% With the current algorithm attempting to find simply the number of objects that would have been detected and classified as galaxies with uniform survey properties, one would expect the stellar structure to leave some imprint on the corrected galaxy map.
% However, this signal should be insignificant due to the much higher counts of true galaxies.
% Therefore, one would hope to see the galaxy map become more uniform, especially in areas where non-uniformity is clearly being caused by artificial variations, such as the $30^{\circ}$ line and the border of the DES footprint.

% Figure \ref{fig:Galaxy_Correction} shows these galaxy corrections.
% Prominent depth features are visually less pronounced, including the border of the footprint.
% Both original and corrected maps have corrected for FracDet variations among pixels, so the border difference is caused by other factors which have been corrected for.

% Kyle: Switch order between stars and galaxies, shorten everything about galaxies, move lots of details into the stars paragraph
In Figure \ref{fig:Star_Correction} we show the observed stellar map on the left and our corrected map on the right. 
For all plots in this section we applied a Gaussian smoothing with a kernel of $0.15^{\circ}$.
The Phoenix isochrone was used to select the objects in all plots in this section.
The stellar density map is non-uniform on large scales, with counts increasing near the Galactic plane (along the left and right edges of the footprint) and near the Sagittarius stream located at RA $\sim 30^{\circ}$ and declination  (Dec.) $\sim -5^{\circ}$. 
As well as these physical variations, features correlated with survey properties (such as the RA $\approx30^{\circ}$ vertical stripe and the artifacts at the edge of the footprint) are present in the initial map and mitigated in the corrected map.  
%While these results look promising, more rigorous validation of our methodology is necessary.
%We turn to this next before showing our corrections to Phoenix.
 
Galaxy counts can provide a probe as to how well the algorithm is working since they should be uniform on large scales across the DES footprint.
Figure \ref{fig:Galaxy_Correction} shows initial counts on the left and the corrected map on the right.
As with stars, in the original galaxy map a number of features correlated with survey properties can be seen. 
These include large scale variations in the average number of objects (e.g., lower right of map), vertical stripes at constant RA (e.g., RA $\approx 30^{\circ}$) due to increased depth, and artifacts at the edge of the footprint. 
The corrected map does not exhibit these features and is more uniform, indicating the observational selection function has been mitigated. 

% Our first test performed demonstrates the total corrective power of our algorithm.
% Just say what the effective weight map is, not many details. Corrected/uncorrected. "We construct an effective weight map for stellar objects by corrected/uncorrected."
% Move the plot to DES corrections section
% Cut lots of this. Say we use these bins to sample the large range of impacts of survey properties.
% Say we want to test everything all at once.
% Use DES to select bins, then retrain.

Before applying these corrections to the Phoenix stellar stream specifically, we validate our methodology (Section \ref{sec:validation}).
For our tests, we construct an effective weight map for stellar objects by taking corrected divided by original counts in Figure \ref{fig:Star_Correction}.
The smoothed effective weight map is shown in Figure \ref{fig:weights} with the boundary removed as it suffers from additional systematics.
Bins of this map are used to sample a large range of impacts on stars from survey properties.

\section{Validation of Methodology}
\label{sec:validation}
This section tests the accuracy of the probabilities calculated in Section \ref{sec:methods}.
We test the total corrective power of our relative detection rates (Section \ref{corr_power}), how this changes with the number of objects used to train (Section \ref{conv_counts}), and how consistent our final corrections are as a function of the number of objects used for training (Section \ref{repeatability}).

\subsection{Overall Corrective Power} \label{corr_power}

\begin{figure*}
    \centering
     \includegraphics[width=0.90\textwidth]{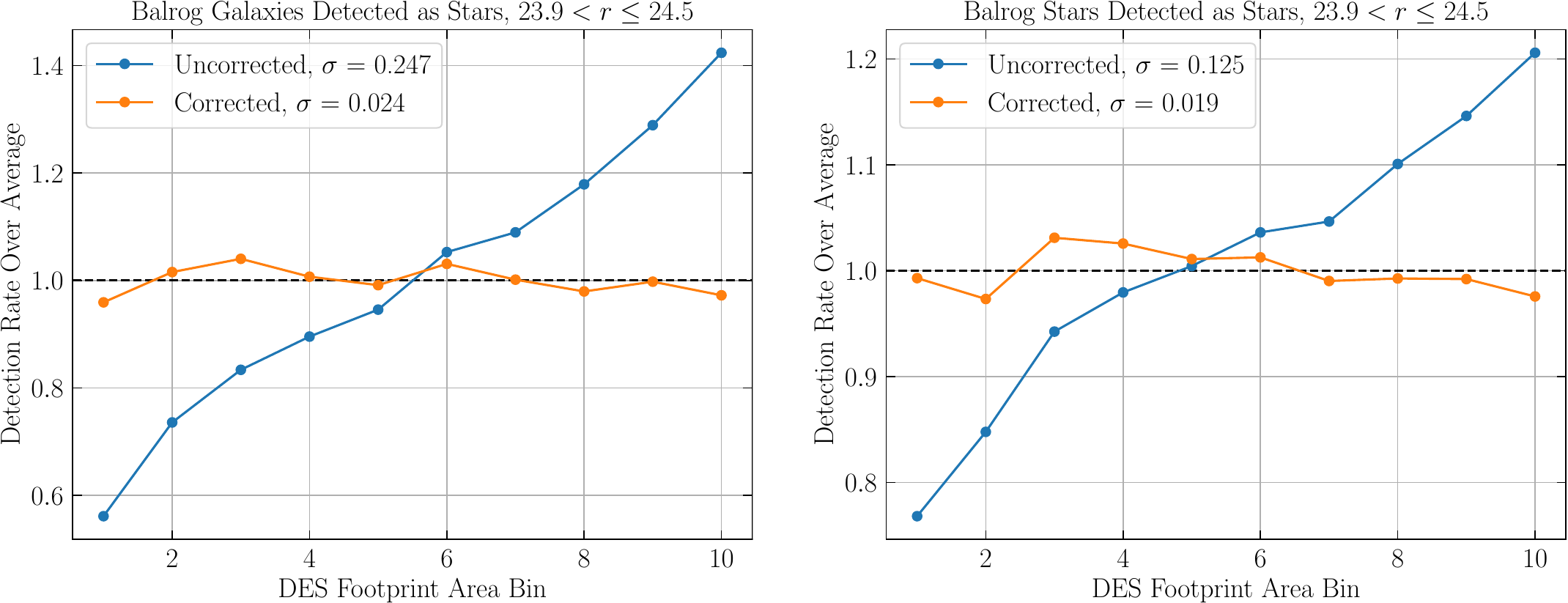}
      \caption{Detection rates relative to the average for a 20\% testing subset of the \texttt{Balrog} objects before and after corrections by effective weight bin number. 
      Before corrections is shown in blue and after corrections in orange. The plot titles describe which relative detection rates are being shown. A drop in variance like observed shows that selection effects are being mitigated.} 
      \label{fig:rel_det}
\end{figure*}

Initially we test the calculations of $D_R\left( C_S,O_S\right)$ and $D_R\left( C_S,O_G\right)$ in our faintest magnitude bin since these are the two relative detection rates necessary for a stellar correction.
We focus on relative detection rates as they have larger variations than classification probabilities.
Assuming uniform injections, the relative detection rate of a type of object (such as true stars classified as stars) is calculated as the number of detected objects of this type per healpixel divided by the overall average.
If the corrections are effective we would expect this ratio to become independent of the amount of correction applied (effective weight). 

Subsets of the data are created by binning the healpixels based on their effective weight from Figure \ref{fig:weights}, to obtain $10$ bins of equal sized samples.
From each area bin, we select $80\%$ of the \texttt{Balrog} objects to be in our training sample which is used to calculate corrections applied to the remaining test set.
Figure \ref{fig:rel_det} shows relative detection rates on the test set across the $10$ effective weight bins, which are referred to in the figure as the DES footprint area bins.
Comparing rates before (blue) and after (orange) corrections shows that variations in relative detection rates for both stars and galaxies drop by a factor of $\sim 5$ .

\subsection{Convergence from Object Counts}\label{conv_counts}

\begin{figure*}
    \centering
     \includegraphics[width=0.90\textwidth]{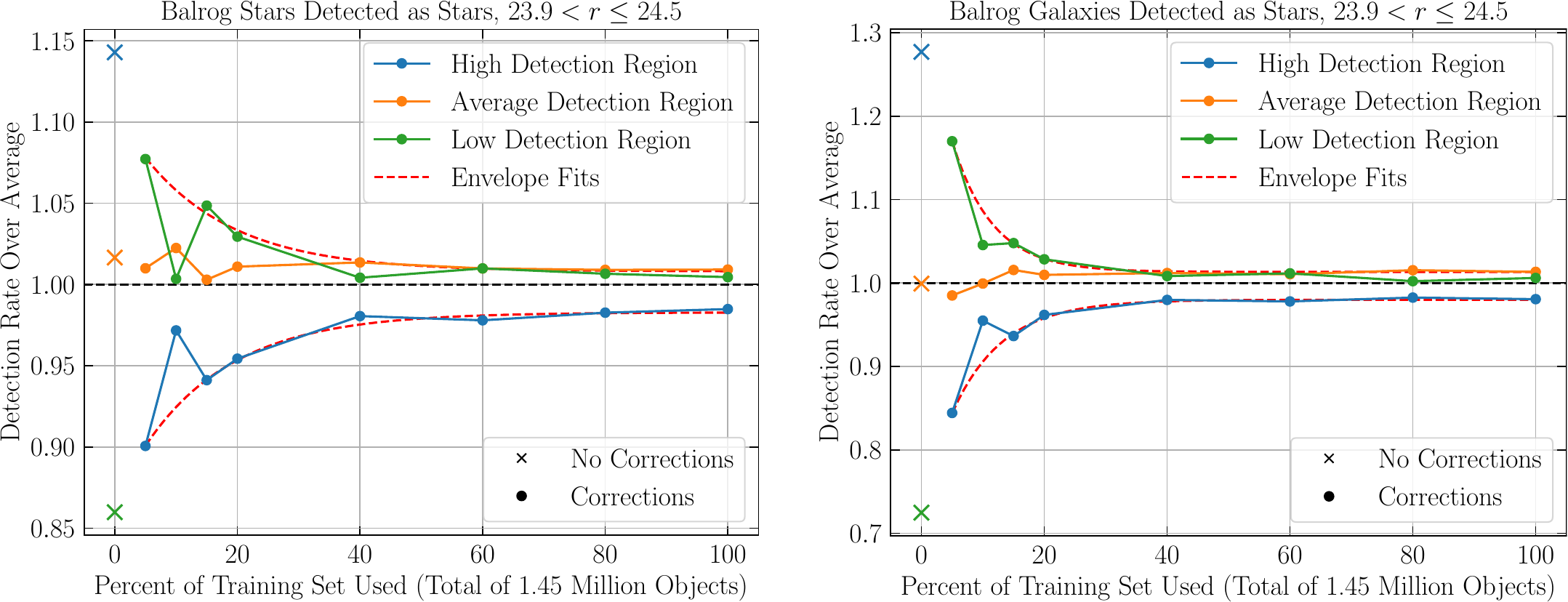}
      \caption{The convergence of corrections as more objects are used for training. A constant 20\% subset of \texttt{Balrog} objects are withheld for testing. The remaining 80\% is then cropped to different levels. The 0\% level indicates that no corrections had been performed. Three area bins are tested. Blue, orange, and green lines represent area bins with above average detection rates, average detection rates, and below average detection rates respectively. Red dotted lines show an exponent fit to an outer envelope. The plot titles describe which relative detection rates are being shown.} 
      \label{fig:three_areas}
\end{figure*}

Characterizing how corrective power scales with more SSI objects will help inform SSI strategies for future analyses.
To investigate this, we perform the test from Section \ref{corr_power} while varying the size of the training sample.
Due to noise concerns, we reduce the number of effective weight bins to 3.
Our results are shown in Figure \ref{fig:three_areas} where we plot the detection rates for correctly classified stars (left) and incorrectly classified galaxies (right) as a function of the size of the training dataset.
Blue (orange, green) points represent an effective weight bin that started with above (approximately, below) average relative detection rates.
To illustrate the general behavior we include an exponential decay envelope in red. 
This line shows convergence is reached with $\sim$40\% of the training set in both cases.
Beyond this point we are limited by the Poisson noise of the test dataset and cannot probe further convergence.

\subsection{Repeatability of Results}\label{repeatability}

\begin{figure*}
    \centering
     \includegraphics[width=0.90\textwidth]{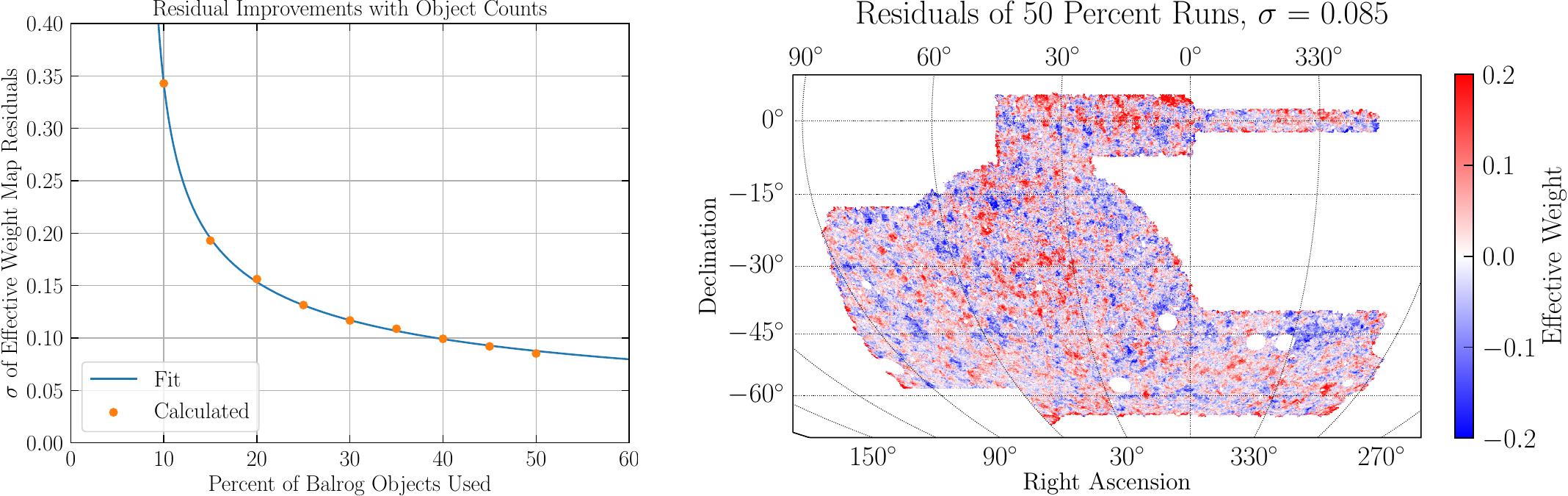}
      \caption{Left: standard deviation of the effective weight map residuals as a function of percentage of training objects used. Right: residuals of the total effective weight map for two runs using 50\% of the \texttt{Balrog} objects.} 
      \label{fig:var_drop}
\end{figure*}

%Another question we would like to answer is how repeatable our results are.
Given a training set, the derived corrections are completely deterministic, but it is interesting to see how these derived corrections change as the training sample is varied. 
This is important because it shows the inherent noise in our corrections as a function of training set size. 
% Our correction algorithm is completely deterministic, %so using the same synthetic sources on two different runs will result in the same corrections.
% but two different realizations with the same number of SSI objects will result in different corrections.
% In this section, we look at how repeatable corrections are as a function of how many objects are injected.

To test this, we vary the objects used to derive corrections. 
At a fixed training set size, we create two disjoint groups of training objects.
For each subset, we derive corrections and effective weight maps. 
We then compute the difference in effective weight, and 
the variance of this residual gives an estimate of our algorithm's repeatability.
This process is repeated for nine training set sizes ranging from 10-50\% of the total sample ($\sim 26$ million total objects). 
Residuals for the two 50\% runs are shown in the right plot of Figure \ref{fig:var_drop}.
% We cannot exceed 50\% since we use two completely disjoint training sets.
The left plot of Figure \ref{fig:var_drop} shows standard deviations of the residuals as a function of the training set size. 
We found our results could be well fit by a power law $y \sim x^{-0.45}$.
%The exact fit is shown on the plot inset.

With \texttt{Balrog} Y6 covering the entire DES footprint instead of just 20\% \citep{Anbajagane:2025:balrog}, we can use our fit to predict the consistency of results when $\sim 5$ times more objects are injected.
In this case, we expect two full runs of \texttt{Balrog} Y6 to have stellar effective weight map residuals with a standard deviation of $\sim 0.03$, compared to $\sim 0.06$ for \texttt{Balrog} Y3.
%Comparing this to \texttt{Balrog} Y3 where our power law fit would predict a standard deviation of 0.06, we can see that the consistency of results from our pipeline should approximately double when \texttt{Balrog} Y6 is used.
Other differences in \texttt{Balrog} Y6 mentioned in \citet{Balrog_Y3} and \citet{Megan_Thesis} such as no longer using delta stars will impact the accuracy of this prediction.

% This can also be used to validate our cycle limit for convergence, lead with this.
Residuals can also be used to validate our cycle limit for convergence as mentioned in Section \ref{sec:methods}.
%Using this residuals approach, we can also address the maximum allowed number of cycles in our iterative training as mentioned in Section \ref{sec:methods}.
We compare our calculated effective weight map to an effective weight map where we put no limit on the number of cycles in the iterative training.
%Two runs were performed for this test.
%In the first run, we limited cycle counts as prescribed in Section \ref{sec:methods} (150 cycles for classification probabilities and 300 for relative detection rates).
%In the second run, we put no limits on cycle counts and let the algorithm keep going until it had achieved the desired accuracy.
The standard deviation for the residuals of these maps was 0.003, showing that the cycle limit we have chosen has allowed results to converge.%is appropriate and the results have converged by the time it is reached.

\section{Application to Stellar Streams}
\label{sec:str_app}
This section applies the corrections from Section \ref{sec:des_app} to observations of stellar streams.
First we look at the Phoenix stellar stream and show that our corrections cause statistically significant linear density shifts (\ref{Phoenix}). % beyond what would be expected from statistical uncertainties.
We then look at synthetic streams and show that our correction algorithm properly suppresses artificial power at length scales larger than the DECam field of view (FOV) (\ref{Synthetic}) of 3 sq. deg.
This suppression of power is detectable on individual stream realizations at length scales relevant for subhalo interactions \citep{DM_Scales}.

% that are predicted by \citet{DM_Scales} to be impacted by subhalos.%dark matter subhalos.

\subsection{Phoenix Stellar Stream}\label{Phoenix}

\begin{figure}
  \includegraphics[width=0.45\textwidth]{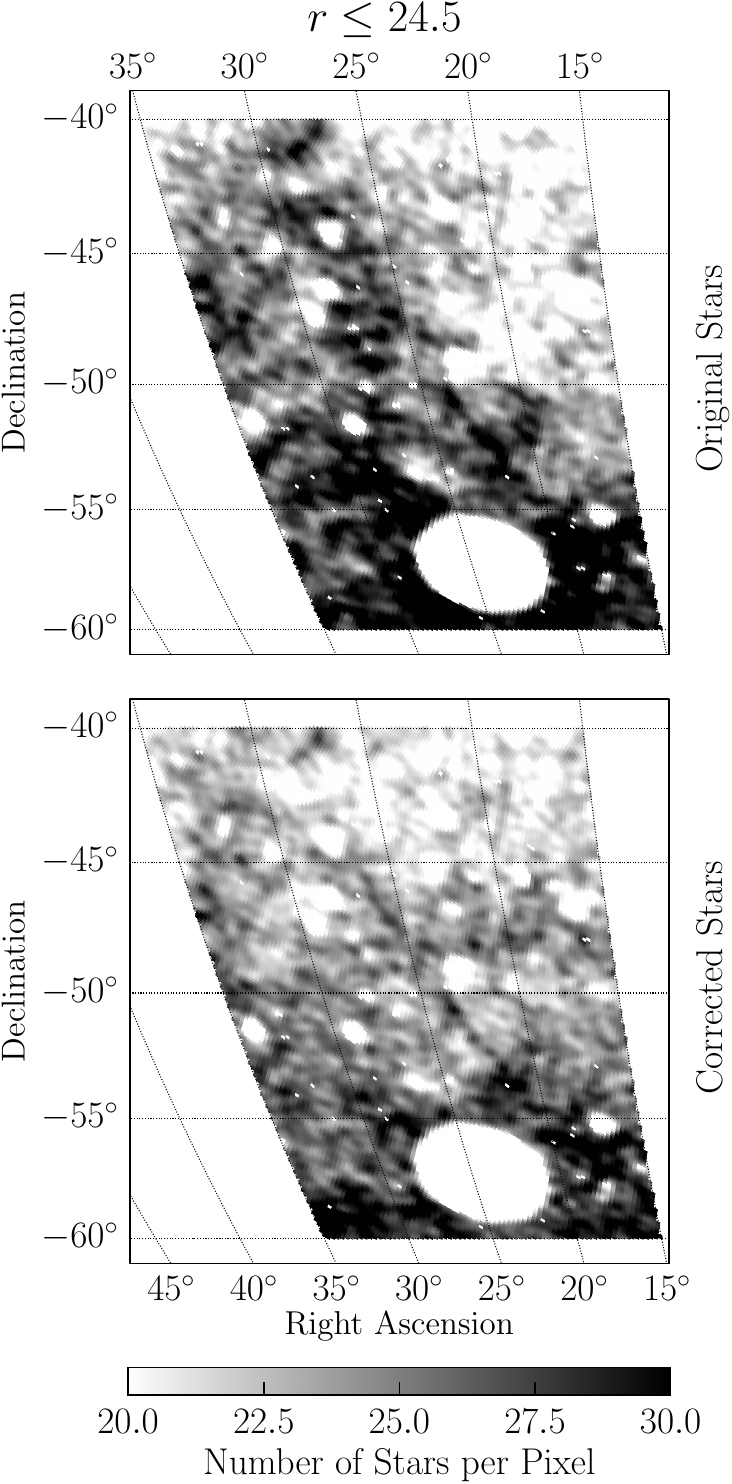}
  \caption{Phoenix stellar stream before and after corrections are applied. A gaussian smoothing with a kernel of $0.15^\circ$ is used to smooth these maps. The (RA, Dec.) endpoints in degrees of this stream are (20.1, -55.3) and (27.9, -42.7) \citep{Phoenix_Endpoints}.}
  \label{fig:Phoenix_Correction}
\end{figure}

\begin{figure}
  \includegraphics[width=0.45\textwidth]{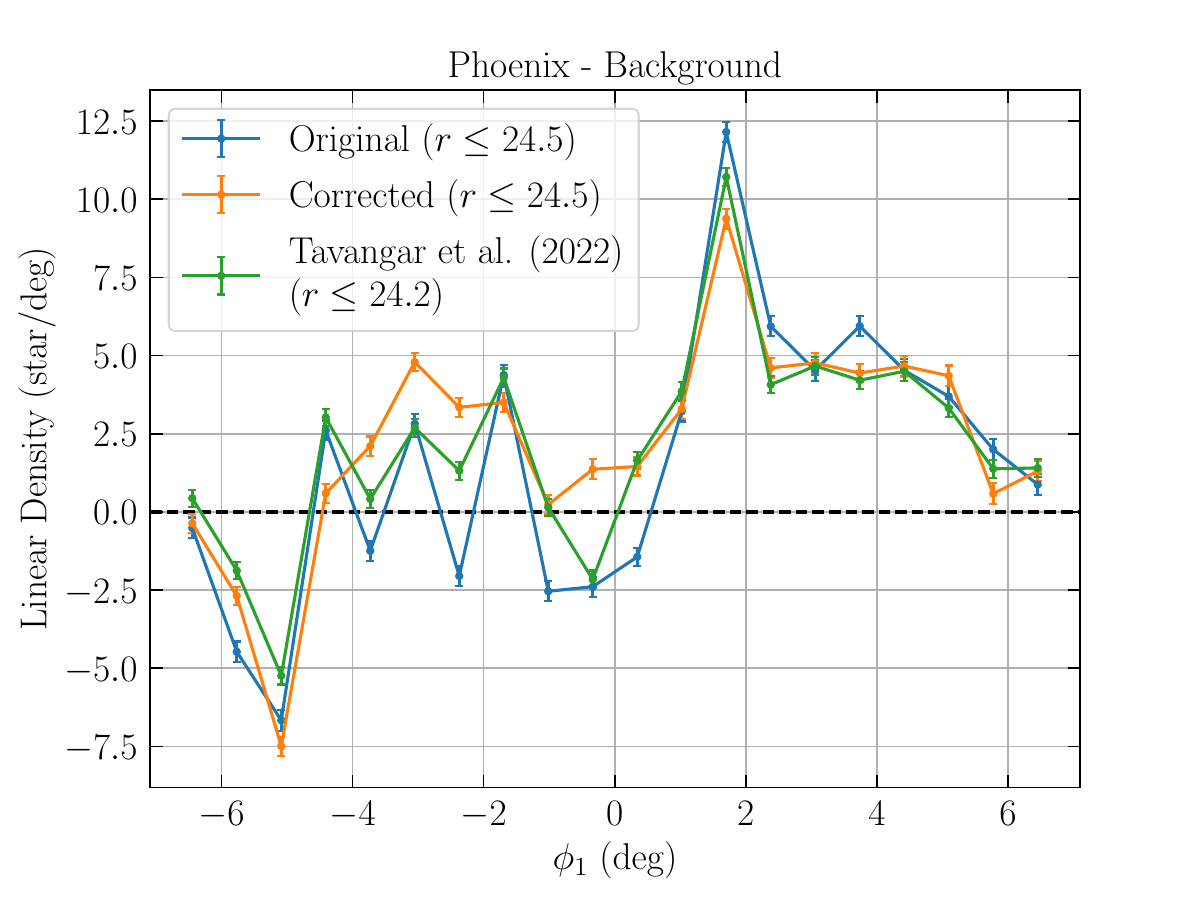}
  \caption{Linear densities of stars per degree of Phoenix minus background. Results are shown for the uncorrected stream, the corrected stream, and the uncorrected stream with the more conservative magnitude cut used in \citet{Phoenix}. Statistical uncertainties on densities are given by errorbars. $\phi_1$ is the angular stream track coordinate. Whether or not these changes are beneficial will be addressed later, for now we just note that the changes are statistically significant.}
  \label{fig:Lin_Dens}
\end{figure}

Using the color and magnitude cuts from Section \ref{matched_filt}, we show the original (top) and corrected (bottom) counts for the on-sky region around the Phoenix stream in Figure \ref{fig:Phoenix_Correction}. % in this Figure.
%Endpoints of the stream are given in the caption. They won't care unless they look at the figure.
It is immediately obvious that the depth feature along RA $\sim 27^\circ$ is removed in the corrected map, enhancing the signal of the stellar stream.
For a quantitative measurement of this effect, we compute the number of excess stars along the stream relative to the background.

%The results of this are shown in Figure \ref{fig:Lin_Dens} where the original 

%, compared to the background. 

% Using the Phoenix stellar stream endpoints and width from Table 1 of \citet{Phoenix_Endpoints} and the rotation matrix defined in Table D.1 of \citet{Stream_Motions} we flag the healpixels within the stream.
% We use this same information to flag background pixels which copy the path and width of the stream but are shifted one degree perpendicular to the stream (in reference to Figure \ref{fig:Phoenix_Correction} our background is to the left of the stream).
We select a background region along the stream but offset by $1^\circ$.
Then we define the linear density as the star counts on stream minus the background. 
The results of this are shown in Figure \ref{fig:Lin_Dens}.
% With star counts on the stream and the background, we look at the linear density of stars per degree of the stream minus the background along the length of the stream, which is shown 
The blue (orange) line signifies the stream before (after) corrections, and the green line signifies the stream with a more conservative ($r\leq 24.2$) magnitude cut from \citet{Phoenix}.
Errorbars represent the statistical uncertainty of each point, showing that the corrections made to the Phoenix stellar stream lead to statistically significant changes in the linear density of the stream.
We note that this is not meant as a direct comparison to \citet{Phoenix} as they use DES Y6 data, but instead is meant to show the dependence of linear density on magnitude cutoffs.

The two most prominent changes as a result of these corrections are the reduced ``gap" centered at $\phi_1 \sim -0.5$ deg and and the fluctuations seen at the local peak centered around $\phi_1 \sim -3$ deg. 
Conversely, the density peak at $\phi_1 \sim 1.5$ deg, hump at $2 < \phi_1 < 6$ deg, and under-density at $\phi_1 \sim -5$ remain generally the same in all cases. 
This analysis also suggests that it would be interesting to apply this correction technique to the deeper Y6 data analyzed in \citet{Phoenix}. 
In the next section, we turn to tests on the power spectra of simulated streams to assess the impact of these corrections.  

% Corrections to the stream make its presence more apparent, largely due to the background being more uniform and the nearby depth feature at RA $\approx 30^{\circ}$ being less prominent.
% To analyze the background, we use Phoenix stellar stream endpoints from Table 1 of \citet{Phoenix_Endpoints} and the rotation matrix defined in Table D.1 of \citet{Stream_Motions} to flag healpixels at distances between $1^{\circ}$ and $2^{\circ}$ perpendicular to the stream track.
% Using this as our background, the standard deviation on background counts before corrections is 6.73 stars per healpixel which drops to 5.81 stars per healpixel after corrections.
% Using the rotation matrix and the width of Phoenix given in Table 1 of \citet{Phoenix_Endpoints} also allows us to isolate the healpixels that contain Phoenix. 
% Comparing stellar counts on these healpixels to the counts nearby in the RA $\approx 30^{\circ}$ region shows us the improved contrast after corrections.
% To stay close to the stream, for the RA $\approx 30^{\circ}$ region we selected healpixels such that $26^{\circ}<$ RA $<29^{\circ}$ and $45^{\circ}<$ DEC $<50^{\circ}$.
% Average counts before corrections in this region were 26.5 stars per healpixel compared to a stream average of 28.0 stars per healpixel.
% After corrections, these counts were 23.1 and 26.1 stars per healpixel respectively, showing that the contrast between the regions approximately doubled.

\subsection{Synthetic Stellar Streams}\label{Synthetic}

% \begin{figure}
%   \includegraphics[width=0.45\textwidth]{figures/ConfidenceIntervalSpectra.pdf}
%   \caption{Power spectra for a uniform injected stream. Shown are the baseline for the true stream (red), the observed stream before corrections (blue), and the observed stream after corrections (orange). Solid lines represent the median power spectra at each point for 5000 realizations, and the 68\% confidence interval is shaded.}
%   \label{fig:power}
% \end{figure}

\begin{figure*}
    \centering
     \includegraphics[width=0.90\textwidth]{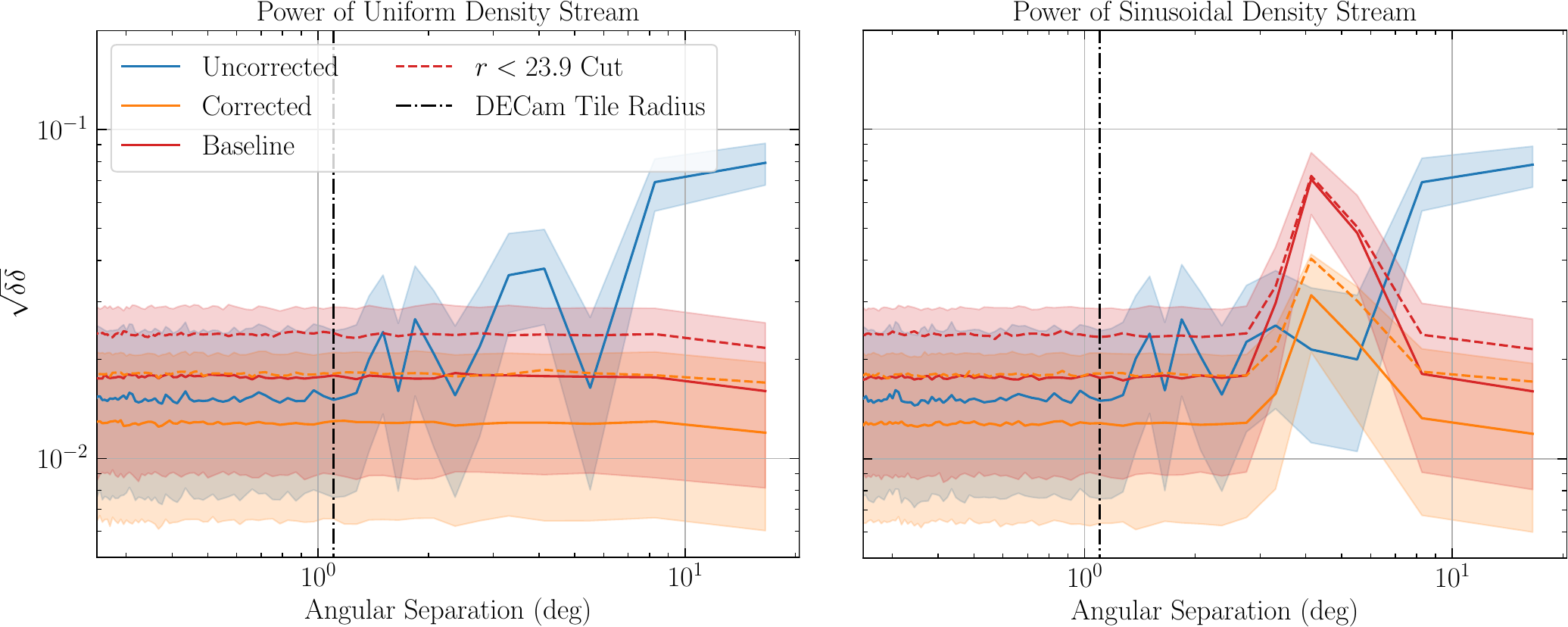}
      \caption{Power spectra for synthetic streams. Shown are the baseline for the true stream (red), the observed stream before corrections (blue), and the observed stream after corrections (orange). We also show the baseline and corrected streams with a more conservative $r<23.9$ cut, denoted with dashed lines. The true stream has perfect classification and no variations in detection rates. Solid lines represent the median power spectra at each point for 5000 realizations. Shaded regions represent the $16$th and $84$th percentiles among the 5000 simulations for each point along the power spectra. Power spectra are shown for a uniform density stream on the left and a sinusoidal density stream, with a period of $4.4^\circ$ (two times the DECam FOV), on the right.} 
      \label{fig:power}
\end{figure*}
% Plot titles: Power of Uniform Density Stream, Power of Sinusoidal Density Stream
To probe the effect of our corrections on the 1D density auto-correlation function, we compute this power spectra for 5000 simulated streams before and after corrections.
These simulated streams are injected into an arbitrary region of the sky at high Galactic latitude with large variations in effective weight (the area is fixed between realizations). 
For each realization, we assume a uniform distribution for the true number of stars and galaxies plus a stream population of stars. 
Since we don't know the absolute detection rate, we fix the number of detected objects for each realization.
We then use relative detection rates to create realistic spatial distributions for all of our mocks.
Our stream is defined to, on average, have a 25\% excess of stars compared to the background stellar population.
Corrections are then performed on the data based on the relative detection rates and correct classification probabilities as prescribed in Section \ref{sec:methods}.
For more details, refer to Appendix \ref{app:adt}.

% Then we 

% first generate realistic counts for mock stars and galaxies.
%We initially inject these objects uniformly, but applying the appropriate detection and classification rates imparts a structure on the background.

%For each realization, we first generate a realistic number of mock stars and galaxies and apply the appropriate detection and classification rates. 
%The generated mock objects include a uniform background and a stream population

%To start our simulations, we selected an area of sky on which to test, which was a $\sim 6^{\circ}$ by $\sim 17^{\circ}$ patch of sky centered at a position of RA $\approx 30^{\circ}$, DEC $\approx 35^{\circ}$.
%Based on our corrections, we estimate appropriate counts for injections of sandbox stars and galaxies.
%From here, we inject stars and galaxies uniformly across the patch of sky.
%Whether the injected objects are detected and what classification they're given is made to align with our calculated detection and classification probabilities (for more details, refer to Appendix \ref{app:adt}).
%An excess 25\% of stars with the same magnitude distribution is injected along a line of dimensions $\sim 0.3^{\circ}$ by $\sim 17^{\circ}$ to act as a stream.

To compare these uncorrected and corrected results to a baseline, we independently do a run with perfect classification and no variations in detection rates.
Two runs were performed, one with a uniform density stream, and one with a sinusoidal density stream with period $4.4^\circ$, two times the DECam FOV.
% This run is indicative of the true underlying stream which the other two runs are trying to estimate. Commented out because this still doesn't have perfect detection, just uniform detection rates. Therefore it's not really the true underlying stream. I think the previous sentence explains enough about what this is.
In Figure \ref{fig:power} we show the results of this exercise.
The lines are the median power spectra for baseline (red), corrected (orange), and uncorrected (blue) results. 
Shaded regions represent the $16$th and $84$th percentiles among the 5000 simulations for each point along the power spectra.
For the uniform density stream (left), the corrected stream obtains a near constant power spectra while the uncorrected stream obtains extra power at scales larger than the DECam FOV.
For the sinusoidal density stream (right), the corrected stream does gain signal power at the expected scales, although its amplitude is lower due to the contamination of galaxies.
The differences from the uncorrected stream can be seen even on individual runs, and as is shown in \citet{DM_Scales}, these are the same angular scales expected to be sensitive to dark matter subhalo interactions.
Dashed lines in Figure \ref{fig:power} represent results with a more conservative magnitude cut of $r<23.9$. 
Noise levels increase as expected, and in the sinusoidal case the signal to noise ratio of the peak is reduced by a factor of $10\%$ compared to the full $r<24.5$ data, showing the advantage of using fainter objects.

%The dashed line represents the FOV radius of the DECam.
%The plot shows the median power spectra as solid lines with a surrounding shaded region representing a 68\% confidence interval.
%Increased power in the uncorrected stream occurs at scales larger than the DECam FOV radius, and the increase in power can be noticed at statistically significant levels on certain scales even for individual runs.
%As is shown in \citet{DM_Scales}, these length scales are sensitive to dark matter subhalos, making corrections here crucial in making unbiased predictions of the subhalo mass distribution function.
We can estimate the importance of proper corrections at even fainter magnitudes where the ratio of galaxies to stars becomes even larger. 
This is particularly relevant for
future surveys such as the Vera C. Rubin Observatory Legacy Survey of Space and
Time \citep[LSST;][]{LSST}. 
To test this we use the DES deep fields to obtain a realistic distribution of object counts at fainter magnitudes ($24.5<r\leq 25$).
%To get an estimate on the importance of corrections in future deeper surveys such as , another test was done using an additional magnitude bin with 
%For these objects, a simpler blue color cut was done with $0\leq g-r\leq 1$.
%Object densities were approximated by using DES deep field object counts.
Assuming survey properties will have similar impacts as in DES, we can get a lower bound for detection and classification variations by setting them equal to the variations from our previous faintest bin ($23.9<r\leq 24.5$).
Running the same power spectrum test as before, the multiplicative difference between the baseline and uncorrected power spectra increased by a factor of 5 at the largest tested scales, showing that corrections will be even more necessary to fully leverage LSST-like data.

\section{Conclusion}
\label{sec:conclusion}
Accurate stellar stream density measurements offer a promising probe into dark matter substructure in the Milky Way.
While variable selection effects from survey properties can introduce artificial density variations, synthetic source injection can be used to correct for these effects.
In this work we describe and offer a method to correct for artificial density variations that will allow for a more accurate characterisation of stellar streams and their parameters.
%In this paper, we have presented a method of using synthetic source injection to correct for these systematic effects. 

% Newly commented out
% Section \ref{sec:methods} begins with a description of our method for calculating correct classification probabilities and relative detection rates. 
% % Briefly, the uniform injection of synthetic objects allows us to directly probe variations in correct classification rates and detection rates, as a function of survey properties. 
% These results can then be used to mitigate the impact of variations in survey properties in both stellar and galaxy density maps.
% The algorithm we use for applying these corrections is also outlined in Section \ref{sec:methods}.

Section \ref{sec:des_app} shows our corrections applied to DES DR1 data. 
After these corrections were applied, galaxy counts became more uniform on large scales.
Features correlated with survey properties such as the depth feature at RA $\sim 30^\circ$ and edge artifacts are also visually diminished after corrections.

Validation of our algorithm is performed in Section \ref{sec:validation}.
%In particular, we characterized the overall corrective power, the behaviour as a function of the number of training objects, and the repeatability of our algorithm as a whole. 
For overall corrective power, we find that key relative detection rates in our faintest magnitude bin had their standard deviations drop by a factor of at least five after corrections.
Running this same test with a variable training set size showed that relative detection rates improved in uniformity when more training objects were used, but only up to a limit.
Finally, we find that the repeatability of our results improved as more training objects were used.
These improvements persisted over the entire range of training set sizes that we were able to test.

We apply our corrections to stellar streams in Section \ref{sec:str_app}.
Here we find that the changes in the linear density of Phoenix due to our corrections are statistically significant, which calls into question some of the apparent density variations reported in the literature.
% We then looked at simulated stream power spectra.
For uniform density simulated streams, our corrections mitigated all signals from survey property variations.
For sinusoidal density simulated streams, our corrections saw a signal at the proper frequency, although the amplitude of the corrected signal was reduced.

One clear next step is to derive corrections for the DES Y6 run of \texttt{Balrog} \citep{Anbajagane:2025:balrog}.
This SSI run has injections across the entire DES footprint compared to the 20\% coverage for \texttt{Balrog} Y3.
One large advantage of this will come in the form of testing set size for the tests performed in Section \ref{sec:validation}.
The number of sources in a 20\% testing set of the \texttt{Balrog} Y6 data would be on the order of the entire \texttt{Balrog} Y3 catalog.
This will make \texttt{Balrog} Y6 less susceptible to potential bias in estimates of relative detection rates that are more likely to arise with smaller sample sizes.
This could lead to the relative detection rates in Figure \ref{fig:three_areas} moving closer to one in the future if the current limits on accuracy were due to biased testing sets.
Additionally, this framework could easily be applied to data from the DECam Local Volume Exploration survey (DELVE; \citealt{Drlica-Wagner:2021}) processed through the DECam All Data Everywhere (DECADE; \citealt{Anbajagane:2025:Decade}) campaign. This data set uses the same instrument with the same data reduction pipeline, but contains much more heterogeneous survey properties.%could easily be applied to the Delve \cite{Drlica-Wagner:2021}/ DECADE \cite{Anbajagane:2025:Decade} dataset which uses data from the same instrument with the same data reduction pipeline, but contains much more heterogeneous survey properties. 

An interesting problem to investigate is how well these corrections transfer to other color-magnitude regions.
Our color-magnitude cuts were specific to the isochrone of Phoenix.
If possible, less restrictive color cuts would allow one training session to potentially provide corrections for multiple stellar streams with different isochrones, and allow for more objects to be used in training.
The potential cost would come from how well the \texttt{Balrog} sources would actually model the stars in the stellar stream if they're in a different area of color-magnitude space.

% As a last potential project, one could attempt to make a correction algorithm to obtain true star counts.
% One large issue with this currently is that finding pure detection probabilities is made difficult by the fact that undetected \texttt{Balrog} objects do not have measured magnitudes.
% A potential fix for this would be to use detected \texttt{Balrog} sources to figure out a map from measured magnitude to true magnitude as a function of survey properties and measured magnitudes.
% This could be applied to GOLD objects so that both training and correcting could be done in true magnitude space, which would allow for the calculation of pure detection probabilities.

This work demonstrates that the use of synthetic sources to correct for variable selection effects is a viable approach.
It also provides a groundwork methodology for applying these corrections in future surveys such as the LSST, which will cover the DES footprint (and extend beyond) and use a similar photometric system.
Rudimentary tests show that we can expect the impacts of survey properties to increase by a factor of at least 5 when fainter magnitude bins are used in LSST. Therefore, accurate corrections and a deep understanding of the observational selection function will be even more important in making unbiased measurements of density variations in stellar streams.

%All code used to make the corrections in this paper is publicly available at the following location:

%\url{https://github.com/Kyle-Boone/ssi_corrections_des_y3_balrog}

\section*{Acknowledgments}
%Uncomment the line below this in the tex file to add a million required acknowledgements
PSF acknowledges support from the DIRAC Institute in
the Department of Astronomy at the University of Washington.
The DIRAC Institute is supported through generous gifts from the Charles and Lisa Simonyi Fund for Arts and Sciences, and the Washington Research Foundation.

Contribution statement: 
KB conducted the analysis, created all plots and tables presented here and led the writing.
PSF provided direct supervision of the research, guidance on the analysis and helped with writing.
MT provided technical expertise in accessing and correctly using the SSI data products. 
KBechtol guided major analysis and content decisions as well as co-supervising the research.
ADW and CMV internally reviewed the paper. 
TYC and BMP provided valuable comments that improved the paper's clarity and quality. 
The remaining authors contributed to this work through the construction of DECam and other aspects of data collection; data processing and calibration; developing widely used methods, codes, and simulations; running pipelines and validation tests; and promoting the science analysis.

Funding for the DES Projects has been provided by the U.S. Department of Energy, the U.S. National Science Foundation, the Ministry of Science and Education of Spain, 
the Science and Technology Facilities Council of the United Kingdom, the Higher Education Funding Council for England, the National Center for Supercomputing 
Applications at the University of Illinois at Urbana-Champaign, the Kavli Institute of Cosmological Physics at the University of Chicago, 
the Center for Cosmology and Astro-Particle Physics at the Ohio State University,
the Mitchell Institute for Fundamental Physics and Astronomy at Texas A\&M University, Financiadora de Estudos e Projetos, 
Funda{\c c}{\~a}o Carlos Chagas Filho de Amparo {\`a} Pesquisa do Estado do Rio de Janeiro, Conselho Nacional de Desenvolvimento Cient{\'i}fico e Tecnol{\'o}gico and 
the Minist{\'e}rio da Ci{\^e}ncia, Tecnologia e Inova{\c c}{\~a}o, the Deutsche Forschungsgemeinschaft and the Collaborating Institutions in the Dark Energy Survey. 

The Collaborating Institutions are Argonne National Laboratory, the University of California at Santa Cruz, the University of Cambridge, Centro de Investigaciones Energ{\'e}ticas, 
Medioambientales y Tecnol{\'o}gicas-Madrid, the University of Chicago, University College London, the DES-Brazil Consortium, the University of Edinburgh, 
the Eidgen{\"o}ssische Technische Hochschule (ETH) Z{\"u}rich, 
Fermi National Accelerator Laboratory, the University of Illinois at Urbana-Champaign, the Institut de Ci{\`e}ncies de l'Espai (IEEC/CSIC), 
the Institut de F{\'i}sica d'Altes Energies, Lawrence Berkeley National Laboratory, the Ludwig-Maximilians Universit{\"a}t M{\"u}nchen and the associated Excellence Cluster Universe, 
the University of Michigan, NSF NOIRLab, the University of Nottingham, The Ohio State University, the University of Pennsylvania, the University of Portsmouth, 
SLAC National Accelerator Laboratory, Stanford University, the University of Sussex, Texas A\&M University, and the OzDES Membership Consortium.

Based in part on observations at NSF Cerro Tololo Inter-American Observatory at NSF NOIRLab (NOIRLab Prop. ID 2012B-0001; PI: J. Frieman), which is managed by the Association of Universities for Research in Astronomy (AURA) under a cooperative agreement with the National Science Foundation.

The DES data management system is supported by the National Science Foundation under Grant Numbers AST-1138766 and AST-1536171.
The DES participants from Spanish institutions are partially supported by MICINN under grants PID2021-123012, PID2021-128989 PID2022-141079, SEV-2016-0588, CEX2020-001058-M and CEX2020-001007-S, some of which include ERDF funds from the European Union. IFAE is partially funded by the CERCA program of the Generalitat de Catalunya.

We  acknowledge support from the Brazilian Instituto Nacional de Ci\^encia
e Tecnologia (INCT) do e-Universo (CNPq grant 465376/2014-2).

This document was prepared by the DES Collaboration using the resources of the Fermi National Accelerator Laboratory (Fermilab), a U.S. Department of Energy, Office of Science, Office of High Energy Physics HEP User Facility. Fermilab is managed by Fermi Forward Discovery Group, LLC, acting under Contract No. 89243024CSC000002

Some of the results in this paper have been derived using the healpy and HEALPix package

\facilities{Blanco (DECam)}

\software{
\texttt{astropy} \citep[]{astropy,astropy2,astropy3},
\texttt{fitsio}\footnote{\href{https://github.com/esheldon/fitsio}{https://github.com/esheldon/fitsio}},
\texttt{HEALPix} \citep[]{HEALPix},
\texttt{healpy} \citep[]{healpy},
\texttt{healsparse}\footnote{\href{https://github.com/LSSTDESC/healsparse}{https://github.com/LSSTDESC/healsparse}},
\texttt{matplotlib} \citep[]{matplotlib},
\texttt{numpy} \citep[]{numpy},
\texttt{scipy} \citep[]{scipy},
\texttt{skyproj}\footnote{\href{https://github.com/LSSTDESC/skyproj}{https://github.com/LSSTDESC/skyproj}},
 \texttt{ugali} \citep[]{Bechtol_Ugali,Drlica_Wagner_Ugali}
}

\clearpage

\appendix
\numberwithin{figure}{section}
\numberwithin{table}{section}

\section{Calculating Classification Probabilities and Relative Detection Rates}
\label{app:prob}
This section will discuss the details in calculating classification probabilities and relative detection rates.
The calculations are nearly identical, so more focus will be given to calculating classification probabilities.
The notation used will be the same as was given in Table \ref{tab:notation}.
As additional notation, $SP$ will refer to a general survey property.

\subsection{Classification Probabilities}

This section will focus on calculating $P\left(C_S|O_S\right)$ for some magnitude bin using \texttt{Balrog} data.
The calculation of $P\left(C_G|O_G\right)$ is completely analogous after switching out the \texttt{Balrog} delta star sample for the \texttt{Balrog} galaxies.
For this subsection, valid healpixels will refer to healpixels with valid values for each survey property which also have a detected \texttt{Balrog} delta star within the magnitude bin of interest.

After applying measured magnitude cuts, two numbers are stored for each valid healpixel.
First is the number of detected \texttt{Balrog} delta stars in the healpixel, $IO_S$.
Second is the number of correctly classified \texttt{Balrog} delta stars in the healpixel, $IC_S \cap IO_S$.
Both of these counts are subjected to quality and magnitude cuts.
The sum over all valid healpixels of $IC_S \cap IO_S$ divided by the sum of $IO_S$ can be thought of as the average correct classification probability:

\begin{equation}
    \label{eq:ave_classification}
    \langle P\left(IC_S|IO_S\right)\rangle = \frac{\sum IC_S \cap IO_S}{\sum IO_S}
\end{equation}

Next we compute the relation between each survey property and the correct classification probability.
For each survey property, the valid healpixels are binned according to the survey property.
For this analysis, we choose to use 10 bins.
In each bin the average survey property value is calculated in each bin as well as the relative correct classification rate compared to the full DES footprint.
These are calculated using Eqs.\ (\ref{eq:ave_sp}) and (\ref{eq:rel_class}) respectively. 
Here, summing over a bin means taking the sum over the healpixels within the bin.

\begin{equation}
\label{eq:ave_sp}
    \langle SP\rangle_{\text{Bin}} = \frac{\sum_{\text{Bin}} SP}{\sum_{\text{Bin}} 1}
\end{equation}

\begin{equation}
\label{eq:rel_class}
    \frac{\langle P\left(IC_S|IO_S\right)\rangle_{\text{Bin}}}{\langle P\left(IC_S|IO_S\right)\rangle} = \frac{\left(\sum_{\text{Bin}} IC_S \cap IO_S\right) / \left(\sum_{\text{Bin}} IO_S\right)}{\langle P\left(IC_S|IO_S\right)\rangle} 
\end{equation}

These ten ordered pairs show the impact the survey property has on classification probabilities.
%The blue line in Figure \ref{fig:classification_variations} shows an example of this dependency for effective exposure time sum in the $i$-band.
With this dependency in mind for each survey property individually, we have to construct the classification probability as a function of every survey property.
This is done in an iterative process.
First we select whichever survey property causes the most variance in $P\left(IC_S|IO_S\right)$ among its bins.
For the $\text{n}^\text{th}$ iteration, we will refer to this survey property as $SP_\text{Max,n}$.
Once this survey property is chosen, we correct for the dependency it causes.
This is repeated iteratively until a termination condition is met.
Notably, there is nothing preventing a correction for one survey property introducing a dependency in another survey property that has already been corrected for.
Due to this, the same survey property could be corrected for multiple times during the training procedure.
For example, we could very well have $SP_\text{Max,1} = SP_\text{Max,50}$.

To begin training, we want to figure out which survey property to correct for first.
% by figuring out which survey property has the largest impact on classification probabilities.
Eq.\ (\ref{eq:rel_class}) is calculated for each bin of each survey property and we use a least squares deviation from one to determine which survey property has the largest impact on classification rates.
This is shown in expression (\ref{eq:class_metric}) which is calculated for each survey property.

\begin{equation}
\label{eq:class_metric}
    \sum_{\text{Bin} = 1}^{10} \left(\frac{\langle P\left(IC_S|IO_S\right)\rangle_{\text{Bin}}}{\langle P\left(IC_S|IO_S\right)\rangle} - 1\right)^2
\end{equation}

Using the notation defined earlier, we will let $SP_\text{Max,1}$ be whichever survey property has the largest value for expression (\ref{eq:class_metric}).
With this survey property chosen, we can use the ten ordered pairs defined in eqs.\ (\ref{eq:ave_sp}) and (\ref{eq:rel_class}) to construct a function from $SP_\text{Max,1}$ values to a relative classification rate.
To achieve this we use a linear interpolation function $f_1$ defined such that the following holds for each bin:

\begin{equation}
    \label{eq:inter_func}
    f_1\left( \langle SP_\text{Max,1}\rangle_{\text{Bin}}\right) = \frac{\langle P\left(IC_S|IO_S\right)\rangle_{\text{Bin}}}{\langle P\left(IC_S|IO_S\right)\rangle}
\end{equation}
We note that a constant value is used for extrapolation.

To correct the \texttt{Balrog} data for the dependency on $SP_\text{Max,1}$ we analyze $f_1\left( SP_\text{Max,1}\right)$ for each valid healpixel with inputs of the $ SP_\text{Max,1}$ values on each healpixel.
The output of $f_1$ can be interpreted as how likely a star is to be classified correctly, for a given healpixel relative to the average correct classification rate.
Therefore to remove the $SP_\text{Max,1}$ dependency, at each valid healpixel the number of correct classifications is divided by the output of the interpolation function.
This is shown explicitly in Eq.\ (\ref{eq:class_correction}), where each value shown is assumed to be the value on one specific healpixel.

\begin{equation}
\label{eq:class_correction}
    \left( IC_S \cap IO_S\right)_\text{New} = \frac{\left( IC_S \cap IO_S\right)_\text{Old}}{f_1\left( SP_\text{Max,1}\right)}
\end{equation}

Substituting these new values on each valid healpixel to Eq.\ (\ref{eq:ave_classification}), we can start this process over to make another correction for a survey property dependency.
This process is repeated until values from Eq.\ (\ref{eq:rel_class}) falls between $0.99$ and $1.01$ for each bin for each survey property.
For notation purposes, let the survey property and interpolation function used on the $\text{n}^\text{th}$ iteration be denoted as $SP_\text{Max,n}$ and $f_\text{n}$ respectively.
As mentioned briefly in Section 2.2, training is stopped after 150 cycles even if convergence has not been reached to avoid over-fitting data.

With training completed, $P\left(IC_S|IO_S\right)$ must now be calculated for all healpixels with valid survey properties so it can be applied to data.
This is done in an iterative process.
To start, we assign each of these healpixels the average classification rate, shown in Eq.\ (\ref{eq:init_class}).
We note that $\langle P\left(IC_S|IO_S\right)\rangle$ in this case will refer to the initial value before any corrections were made to $IC_S\cap IO_S$ which could change the value slightly.
In practice, the specific starting point is not important as the average is re calibrated (see 4.2).

\begin{equation}
    \label{eq:init_class}
    P_0\left(IC_S|IO_S\right) = \langle P\left(IC_S|IO_S\right)\rangle
\end{equation}

After this, for each correction to $IC_S \cap IO_S$ applied during training in Eq.\ (\ref{eq:class_correction}), we must apply a corresponding correction to our classification rate.
For concreteness, suppose there were a total of $\text{N}$ cycles used in training.
Then, for $\text{n}$ between $1$ and $\text{N}$ we define the recursive relation given in Eq.\ (\ref{eq:iter_class}), which is calculated on each healpixel.
Once all iterations have been completed, we have our overall correct classification probability. 
Combining that with the relation in Eq.\ (\ref{eq:prob_match}) gives us Eq.\ (\ref{eq:final_class}).

\begin{equation}
    \label{eq:iter_class}
    P_\text{n}\left(IC_S|IO_S\right) = P_{\text{n} - 1}\left(IC_S|IO_S\right) f_\text{n}\left( SP_\text{Max,n}\right)
\end{equation}

\begin{equation}
    \label{eq:final_class}
    P\left(C_S|O_S\right) = P_\text{0}\left(IC_S|IO_S\right) \Pi_{\text{n} = 1}^{\text{N}} f_\text{n}\left( SP_\text{Max,n}\right)
\end{equation}

To avoid a non physical probability, this is cropped to be less than $1$ (negative values will not appear by construction).

\subsection{Relative Detection Rates}

In this section we will calculate four relative detection rates as functions of survey properties for some particular magnitude bin: the rates at which \texttt{Balrog} delta stars are classified as stars $D_R\left( IC_S,IO_S\right)$, \texttt{Balrog} delta stars are classified as galaxies $D_R\left( IC_G,IO_S\right)$, \texttt{Balrog} galaxies are classified as stars $D_R\left( IC_S,IO_G\right)$, and \texttt{Balrog} galaxies are classified as galaxies $D_R\left( IC_G,IO_G\right)$.
For now we focus specifically on $D_R\left( IC_S,IO_S\right)$ but the same process is repeated for the other relative detection rates. 
Valid healpixels in this section will refer to healpixels with valid values for each survey property which also had a \texttt{Balrog} delta star injected which passed quality cuts.%have an injected \texttt{Balrog} delta star of any magnitude.

Compared to the classification rates of the previous subsection, with relative detection rates we are unable to get an average detection rate for our objects.
This is due to the fact that undetected objects do not have measured magnitudes to bin based on.
Therefore, we are restricted to calculating relative detection rates instead of actual detection probabilities.

The calculation of these rates is performed in a similar manner as in Section 5.3.1 of \citet{Galaxy_Corrections}.
Within a bin from some survey property, we look at the average number of detections of interest (\texttt{Balrog} delta stars classified as stars within our magnitude bin) per healpixel in that bin to find a characteristic detection density.
This is then compared to the average number of detections of interest per healpixel across all of the valid healpixels.
If the distribution of objects that could cause a detection of interest is uniform on large scales, the ratio of these two densities will then represent the rate at which detections of interest occur in the survey property bin relative to the rate at which detections of interest occur overall.
Since \texttt{Balrog} objects are injected on a uniformly spaced hex grid and the distribution of objects on this grid is random, this approach is valid for our calculations.

With this framework in mind, we turn now to the details of the actual calculation.
Like in classification calculations, for each survey property we bin the valid healpixels according to the survey property and calculate the relative detection rate as the following:

\begin{equation}
\label{eq:rel_det}
    \langle D_R\left(IC_S, IO_S\right)\rangle_{\text{Bin}} = \frac{\left(\sum_{\text{Bin}} IC_S \cap IO_S\right) / \left(\sum_{\text{Bin}} 1\right)}{\left(\sum IC_S \cap IO_S\right) / \left(\sum 1\right)}
\end{equation}

Note that the numerator is exactly the average number of detections of interest per healpixel in the bin while the denominator is the average number of detections of interest per healpixel across all valid healpixels.

Treating the value calculated in Eq.\ (\ref{eq:rel_det}) like the relative classification rates of the previous subsection, training is then performed.
As mentioned briefly in section 2.2, training is stopped after 300 cycles even if convergence has not been reached.

Once training is complete, we must calculate $D_R\left( IC_S,IO_S\right)$ on all healpixels with valid survey properties. 
This is also performed in a nearly identical way to the calculations in the previous subsection, with the only change being that we start with the assignment:

\begin{equation}
    \label{eq:init_det}
    D_{R,0}\left( IC_S,IO_S\right) = 1
\end{equation}

This is done since our our relative detection rate should by definition be centered on 1 instead of an average probability like the classification probability.
With this assignment done, the recursive relation used to build up to $D_R\left( IC_S,IO_S\right)$ is defined in an identical way as in Eq.\ (\ref{eq:iter_class}).
As a relative rate can exceed $1$, the crop done in the previous section is unnecessary here.

\section{Testing Separability Assumption}
\label{app:sep}
\begin{figure*}
\centering
  \includegraphics[width=0.90\textwidth]{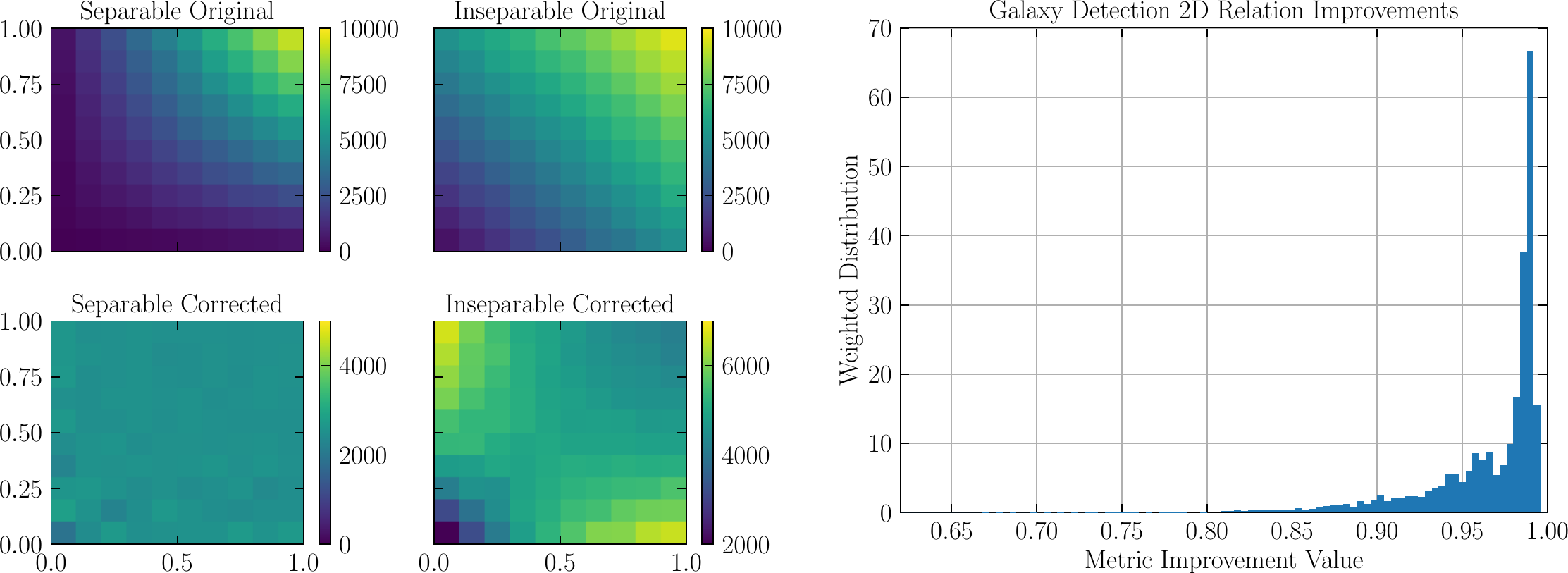}
  \caption{(Left) Example of 2D distributions of objects in inseparable and separable cases before and after corrections. (Right) A weighted distribution of the metric improvement value, weighted by the pre-correction metric. Any value over zero is considered improvement, and a one would be perfect removal of any dependency. Every metric improvement value was greater than $0.65$, with the vast majority being above $0.95$.}
  \label{fig:Sep_Metric}
\end{figure*}

\begin{figure*}
    \centering
     \includegraphics[width=0.90\textwidth]{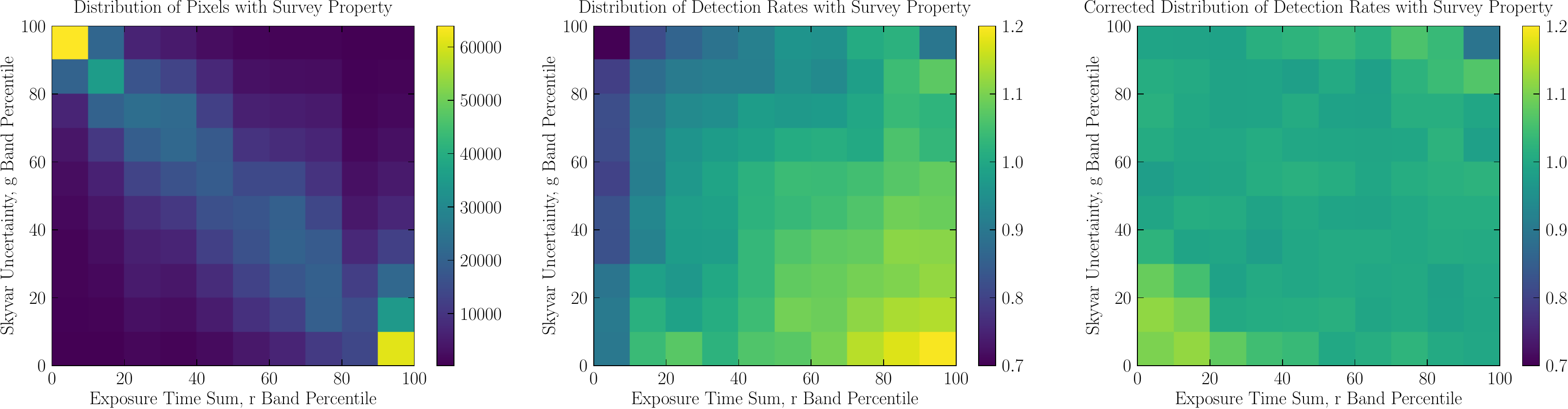}
      \caption{Above is a sample 2D relation using exposure time sum in the r band and skyvariance uncertainty in the g band. For the test 20\% of \texttt{Balrog} objects, first we plot out the healpixel distribution in these two dimensions, shown on the left. With this healpixel distribution, we can use original and corrected detection counts to look at detection rate variations before corrections (in the middle) and after corrections (on the right).} 
      \label{fig:2D_Galaxy}
\end{figure*}

%Mention where this assumption comes up and briefly describe the process through which these higher dimensional dependencies can be hidden.
When calculating relative detection rate and classification probability variations, our iterative approach means that the selection function we learn is separable.
If the selection function is not separable, we can still get convergence in training without properly learning the selection function.
This could occur since convergence is defined by looking at the projections of the selection function onto each survey property individually.
Higher dimensional dependencies could be hidden from view when these projections occur, which would lead to converging to the wrong selection function.
To demonstrate this, simulations were run with injecting objects uniformly over two survey properties, $x$ and $y$.
In one simulation, detection rates were given by a separable function of $x$ and $y$, while in another the detection rate function was inseparable.
Results before and after corrections are shown in the left of Figure \ref{fig:Sep_Metric}, showing the higher dimensional structure present post corrections in the inseparable case.

%Introduce test and train data sets used for this test and why it was used (largest sample of objects).
To test whether this occurred in our pipeline, we used the largest set of \texttt{Balrog} objects possible: all \texttt{Balrog} galaxy injections.
To keep the statistical significance as high as possible, we trained and corrected for overall relative detection rates (being classified as either a star or a galaxy).
Training was done on an 80\% subset of objects, with testing being done on the remaining 20\%.
For each combination of two survey properties we looked at the distribution of valid healpixels over the two survey properties, and the relative detection rates before and after corrections over the two survey properties.
%After corrections were performed, for each combination of two survey properties we looked at three different two dimensional histograms: the overall valid pixel counts in each bin of the two survey properties, and the relative detection rates before and after corrections for each of the bins.
An example of this is shown in Figure \ref{fig:2D_Galaxy}.
Ideal results would be uniform distributions in the corrected plot, especially in regions with high healpixel counts as these are the most common throughout the footprint.

With this example shown, we next designed a metric to concisely summarize results over every combination of survey properties.
Overall, both the original and corrected two dimensional histograms should have values centered about one. 
Furthermore, the points we care about more are the points with high healpixel densities, as this will impact a greater portion of the DES footprint.
To take these factors into account, we use a metric of variance from one weighted by healpixel count. 
If we let $i$ index over each two dimensional bin in the histograms from Figure \ref{fig:2D_Galaxy}, let $PC_i$ be the healpixel counts in bin $i$, and let $D_{R,i}$ be the detection rate variation in bin $i$ (these are the histogram values of the two rightmost histograms of Figure \ref{fig:2D_Galaxy}), we write our metric as:

\begin{equation}
\label{eq:2d_metric}
\epsilon = \frac{\sum_i PC_i\left(1 - D_{R,i}\right)^2}{\sum_i PC_i}
\end{equation}

This metric will be non-negative by construction, and better results will lead to values closer to zero.
Using these features we then define an improvement metric.
If for any given survey property combination $\epsilon_{\text{orig}}$ is the metric value on the original detection variations and $\epsilon_{\text{corr}}$ is the metric value on the corrected detection variations, we define out improvement metric as:

\begin{equation}
\label{eq:met_impr}
\epsilon_{\text{impr}} = 1 - \frac{\epsilon_{\text{corr}}}{\epsilon_{\text{orig}}}
\end{equation}

Any positive value for $\epsilon_{\text{impr}}$ suggests an improvement, with a value of one being perfect.
We calculated $\epsilon_{\text{impr}}$ values for every combination of survey properties using the 20\% testing data from earlier.
We do not necessarily care about each $\epsilon_{\text{imp}}$ value equally, so weights of $\epsilon_{\text{orig}}$ are also stored for each survey property combination.
If two survey properties hardly had any detection rate dependencies in their 2D relation to begin with their improvement matters far less than two survey properties who initially had high dependencies.
With these weights in mind, a weighted histogram of $\epsilon_{\text{impr}}$ was generated, as is shown in the right plot of Figure \ref{fig:Sep_Metric}.
The overwhelming trend towards 1 shows that our separable assumption is reasonable.

%Describe improvement metric with an intuitive description and show plot of its weighted values which shows the separability assumption holds to a high degree of accuracy.

\section{Relative Detection Rate Correlations}
\label{app:rdrc}
\begin{figure*}
    \centering
     \includegraphics[width=0.90\textwidth]{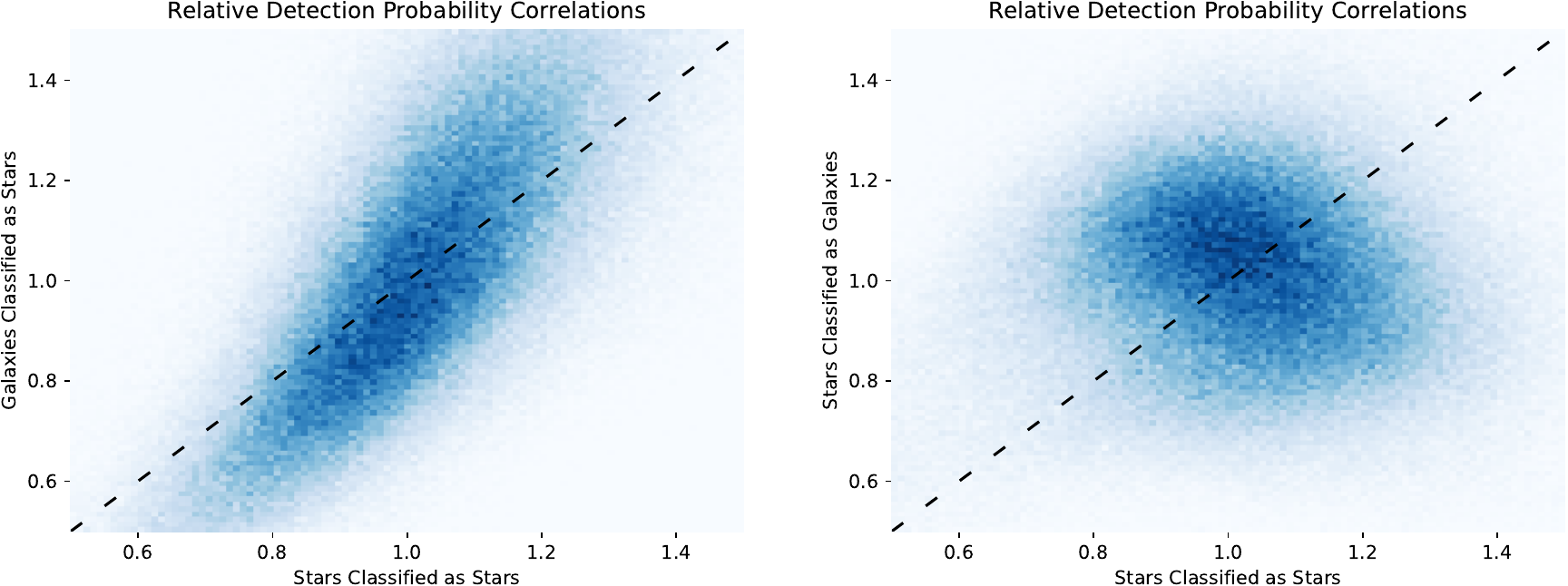}
      \caption{Each plot is using relative detection rates for our faintest magnitude bin. The dotted black line shows where the two rates are equal. (Left) Relative detection rates of stars detected as stars vs galaxies detected as stars. (Right) Relative detection rates of stars detected as stars vs stars detected as galaxies.} 
      \label{fig:correlations}
\end{figure*}

In Figure \ref{fig:correlations} we show the distributions of two different relative detection rate combinations across the DES footprint.
For both plots we focus on relative detection rates in our faintest magnitude bins.
The plot on the left shows the relative detection rates of stars detected as stars vs galaxies detected as stars.
In general this plot shows that the relative detection rate of galaxies classified as stars has a higher variance, and there is a clear positive correlation.
This correlation is likely due to the fact that galaxies classified as stars appear to be point sources and so survey properties that affect detection rates of stars will affect these galaxies as well. 
The plot on the right shows the relative detection rates of stars detected as stars vs stars detected as galaxies.
This plot does not show strong correlations between the two relative detection rates.
The absence of a correlation implies that the detection rate of stars that are then classified as galaxies is not affected by survey properties in the same manner. 
One possible explanation for this could be that unrecognized blends are causing brighter stars to be classified as galaxies, and therefore higher changes in survey properties will not affect the detection rate of these objects.

Overall, the scatter in these relations eliminates the possibility of certain simplifications that could have potentially been made to our correction algorithm.
Referring back to Section \ref{sec:methods}, our current correction algorithm, originally given in Eq.\ (\ref{eq:FCS}), is given by:

\begin{equation}
\label{eq:app_FCS}
    FC_S = \frac{C_S\cap RO_S}{D_R\left( C_S,O_S\right)} + \frac{C_S\cap RO_G}{D_R\left( C_S,O_G\right)}
\end{equation}

If the left plot of Figure \ref{fig:correlations} showed that the two relative detection rates used above were equal, we could simplify that correction to be:

\begin{equation}
\label{eq:app_FCS_simplify}
    FC_S = \frac{C_S}{D_R\left( C_S,O_S\right)} = \frac{C_S}{D_R\left( C_S,O_G\right)}
\end{equation}

If this were the case, there would be no need to calculate classification probabilities since we would not need $RO_S$ or $RO_G$. 
As well as this, we would only need to calculate one of $D_R\left( C_S,O_S\right)$ or $D_R\left( C_S,O_G\right)$.

The correction method that we did not wind up using, originally given in Eq.\ (\ref{eq:FCS2}), is given by:

\begin{equation}
\label{eq:app_FCS2}
    FC_S = \frac{C_S\cap RO_S}{D_R\left( C_S,O_S\right)} + \frac{C_G\cap RO_S}{D_R\left( C_G,O_S\right)}
\end{equation}

If the right plot of Figure \ref{fig:correlations} showed that the two relative detection rates used above were equal, we could simplify this correction in a similar fashion as in Eq.\ (\ref{eq:app_FCS_simplify}).
The scatter in both plots shows that neither correction algorithm can be simplified, so all relative detection rates must be calculated.

\section{Algorithm Design Tests}
\label{app:adt}
%Begin by reminding what the two competing algorithm designs are.
In Section \ref{sec:methods}, we presented two different correction algorithms: one to estimate what the number of objects classified as a stars would have been with uniform survey properties (we will refer to it as the recovered classified algorithm), and one to estimate what the true number of observed stars would have been with uniform survey properties (recovered observed).
The recovered classified algorithm was chosen for use throughout this work.
This section will discuss tests performed on the two algorithms that were used in the making of that decision.

%Mention how sampling will be done for the stream every time, including which area of sky was chosen and a plot of the streams at different points in the pipeline. Edit down what's written below.

Each test we run uses the same modeling sandbox to generate the counts we analyze.
For an area on which to test, we initially chose a $3\times3$ square of 32 resolution healpixels centered at a position of RA$=30^{\circ}$, Dec.$=-35^{\circ}$.
This area was chosen since it contains the depth feature at RA$\approx30^{\circ}$.
This area then has its resolution increased to 512 (giving 2304 total healpixels), since this is the resolution at which we apply our corrections to real data.
A subset of these healpixels (which is test dependent) is chosen to hold the simulated stellar stream.

Applying our correction pipeline to this area of sky with real data gives us estimates for the counts in the area for objects of type $O_S\cap C_S$, $O_G\cap C_S$, $O_S\cap C_G$, and $O_G\cap C_G$ for each magnitude bin we use.
Injections for each of these four object types occur randomly on healpixels, and detection probabilities for injections are made to be proportional to the relative detection rate for the type of object being injected.
This continues until we reach our calculated counts for each type of object.
For our simulated stream, an excess percentage of $O_S\cap C_S$ and $O_S\cap C_G$ type objects are then injected and then cropped to the stellar stream healpixels.
After sampling, $O_S\cap C_S$ and $O_G\cap C_S$ counts are summed to get uncorrected star counts while $O_S\cap C_G$ and $O_G\cap C_G$ counts are summed to get uncorrected galaxy counts.
%With information on relative detection rates, we can then create a simulated realization.
%As parameters, the stream pixels and excess percentage of stars in the stream are required.
%Once this is given, sampling begins.
%When injecting objects of a particular type and in a particular magnitude bin, a random pixel is chosen, and the probability of an object actually being detected is chosen so that the relative probabilities match up with our calculated relative detection rates.
%Injections continue until we hit the quota for each type of object in each magnitude bin.
%For $O_S\cap C_S$ and $O_S\cap C_G$ type objects, this is repeated with the excess percentage times the quota as a new quota.
%This second sampling is then cropped to just the stream pixels and then added to the first sampling.
%One could alternatively imagine multiplying the relative detection rates by some average detection rate and then using expectation values to determine how many objects should be injected to achieve the quota counts as an expectation value of detected counts.
%This would achieve more variation in counts, but it assumes a detection rate, which is something that we cannot calculate at this time.
%In practice, results for the second test we will mention are the only ones impacted by changing which sampling method is used.

\begin{figure*}
    \centering
  \includegraphics[width=0.90\textwidth]{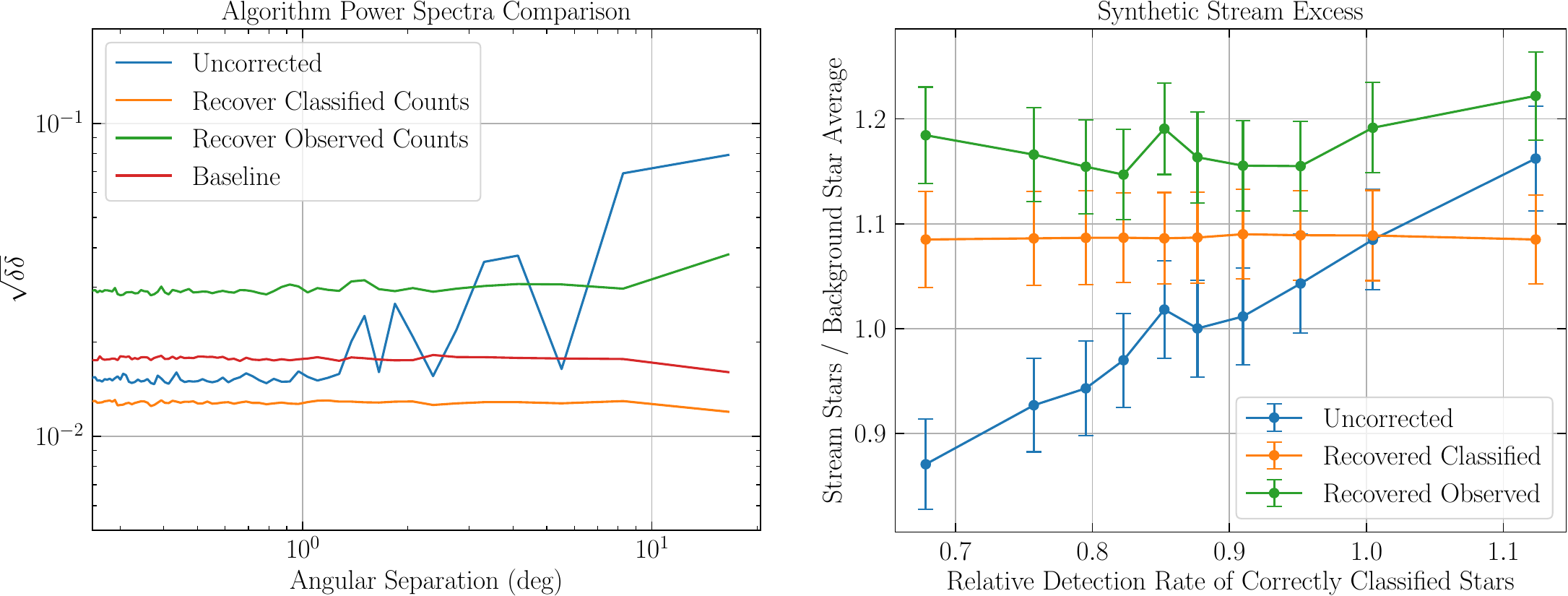}
  \caption{(Left) Median power spectra over 5000 realizations for a baseline stream (red), uncorrected data (blue), and corrected data with each correction algorithm (orange and green). The recover classified counts algorithm shows lower noise levels than the recover observed counts algorithm. (Right) One dimensional dependencies of stream density from the relative detection rate of correctly classified stars in our faintest magnitude bin ($23.9<r\leq 24.5$). 5000 realizations are performed, and the averages are plotted with standard deviation error bars. Both corrected lines show decreased variance. \label{fig:algo_tests}}
\end{figure*}

% Start by describing the power spectrum test since this is the most powerful one.
For our first test, the different correction algorithms were tested on their ability to recover the power spectrum of a stellar stream.
For this test the overall size of the sandbox was increased to a $3\times9$ rectangle of 32 resolution healpixels before being increased to a resolution of 512.
432 healpixels were used for the stream ($3\times144$) to make a stream that was approximately $17^{\circ}$ long, slightly longer than Phoenix.
An excess stream percentage over the background of 25\% was used for each of the 5000 realizations.
As a baseline to compare to, 5000 realizations were done with uniform detection rates and no misclassification. % modeling what the actual underlying stream would look like. Not perfect detection so not really the true underlying stream.
For each of the 5000 realizations, a density power spectrum was calculated.
The median of these power spectra for each algorithm is shown in the left plot of Figure \ref{fig:algo_tests}.
The underlying baseline result is shown in red, uncorrected in blue, recovered classified in orange, and recovered observed in green.
Both correction algorithms remove most structure from the power spectra, but the recovered classified algorithm results in a median power spectrum that has lower noise than the recovered observed algorithm.

% Next, do the 1D dependency tests on relative detection rates to show increased stability but lower accuracy on our model.
The next test we ran looked at dependencies of stellar stream density as a function of the different relative detection rate variations.
For this test, a random 230 of the total 2304 healpixels are chosen to host the stream (the structure of the simulated stream is unimportant for this test).
The excess stream percentage was chosen to be 25\%.
After both correction algorithms are applied, we bin the stellar stream healpixels into 10 bins in terms of each of the 12 relative detection rates used (4 types of relative detection rates in 3 magnitude bins).
With this, we check average star counts in each bin and compare it to the average number of stars in the background.
An ideal result would be a flat curve (showing no detection rate dependency) at a value of 1.25 (since 25\% extra stars were injected).
Results from of this test are displayed in the right plot of Figure \ref{fig:algo_tests}.
In this plot we look at the stream excess over background as a function of the relative detection rate of correctly classified stars in our faintest magnitude bin.
Uncorrected results are shown in blue, recovered classified in orange, and recovered observed in green.
For our selected stream healpixels relative detection rates were lower than average which causes the $x$ axis to be centered below 1.
The recovered classified curve can be seen to have lower variance while the recovered observed curve is closer to 1.25.

%End with the excess stars injected test as just a sanity check that our algorithm is working as it should.

% A final test was done to test the responsiveness of the average stream density to the number of injected stars.
% As with the previous test, a random 230 of the total 2304 healpixels are chosen to host the stream.
% Each of the 5000 realizations had excess stream percentages ranging from 0\% to 100\% of the background density.
% Once corrections were done, we looked at the average number of stars per healpixel within the stream.
% We would expect this number to grow linearly with the percentage of extra stars injected, which is what we found as can be seen in the right plot of Figure \ref{fig:two_algo_tests}.
% As well as this, the horizontal variance from the line of best fit can be viewed as an uncertainty in recovering the true excess given some measurement of the stream.
% For the uncorrected plot, based on our sampling procedure the horizontal variance arises from shot noise.
% Comparing this to the corrected variances can show us if these correction methods are introducing additional noise in obtaining the underlying true excess.
% With this in mind, the horizontal standard deviation around the line of best fit for the uncorrected plot is $0.46$ stars per healpixel, compared to values of $0.46$ and $0.53$ for the recover classified counts and recover observed counts algorithms respectively.

Both of these tests show the increase in noise when using the recovered observed algorithm. 
For our application in particular of looking at stellar stream density fluctuations, the power spectrum test is the most useful since power spectra can be used for dark matter constraints.
In the power spectrum test in particular, the recovered classified algorithm matched the baseline stream to a higher degree than the recovered observed algorithm.
This was our reasoning for using the recovered classified algorithm.
It should be noted that for structured streams, if the power of the structure exceeds the noise from the recovered classified algorithm, then the recovered algorithm gives a more accurate value for the power.
This could motivate a combination of the two algorithms being used in certain applications.

% \begin{figure*}
%     \centering
%   \includegraphics[width=0.90\textwidth]{figures/PowerSpectraComp.pdf}
%   \caption{(Left) Median power spectra over 5000 realizations for a baseline stream, uncorrected data, and corrected data with each correction algorithm. No added structure is given to relative detection rates. (Right) Here structure has been added to relative detection rates at a scale of 4.4 degrees, twice the size of the DECam footprint. \label{fig:power_spectra}}
% \end{figure*}

% \begin{figure*}
%     \centering
%      \includegraphics[width=0.90\textwidth]{figures/LastTwoAlgoTests.pdf}
%       \caption{(Left) One dimensional dependencies of stream density on relative detection rates. This particular plot is for the relative detection rate of stars being classified as stars in the faintest magnitude bin. 5000 realizations are performed, and the averages are plotted with standard deviation error bars. Both corrected algorithms show a decreased dependency. (Right) Average number of stars in a stream healpixel as a function of the excess percentage in the stream (which has been converted to the excess stars per healpixel).} 
%       \label{fig:two_algo_tests}
% \end{figure*}

\section{Determining Magnitude Bins}
\label{app:mag}
%Start by mentioning the magnitude cutoff at a point where detection rates had started to trend towards 0.

\begin{figure*}
    \centering
     \includegraphics[width=0.90\textwidth]{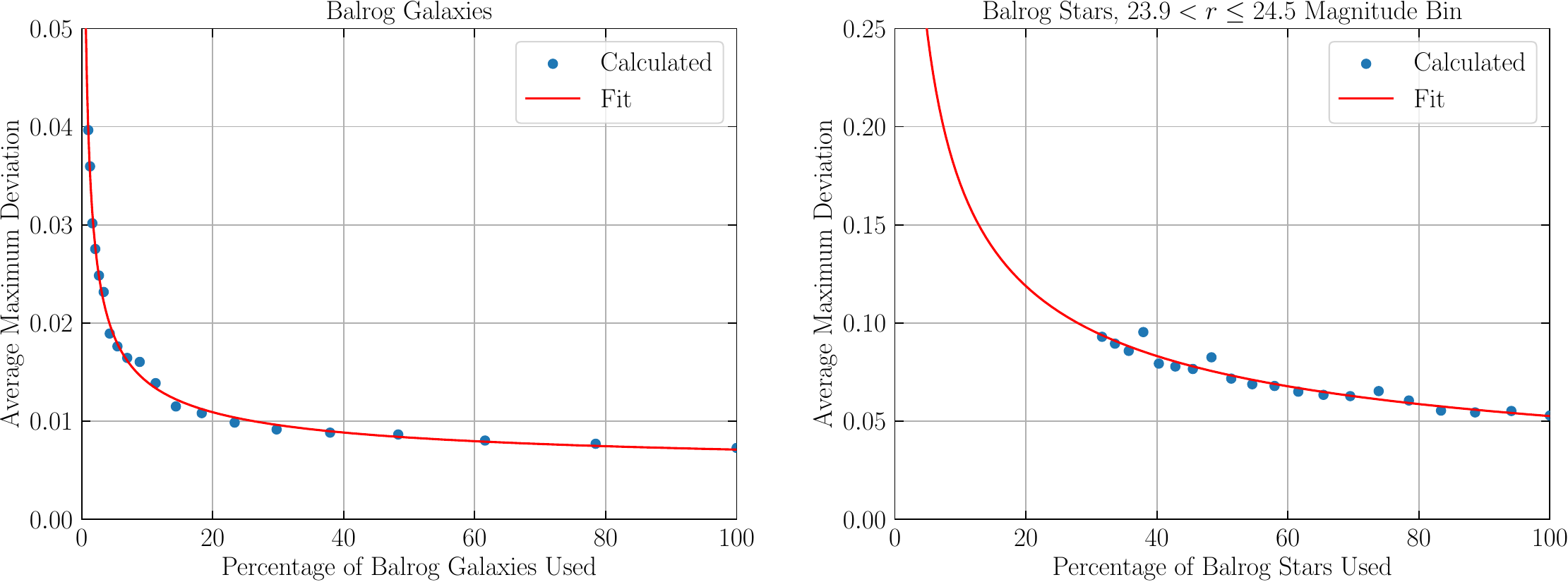}
      \caption{(Left) Average maximum deviations for relative detection rate variations for \texttt{Balrog} galaxies. Higher object counts allowed for probing at lower percentages of the \texttt{Balrog} sample. (Right) Average maximum deviations for relative detection rates variations for \texttt{Balrog} stars classified as stars in the faintest magnitude bin used.} 
      \label{fig:bal_num}
\end{figure*}

Object magnitude plays a role in detection rate and classification probability variations, so to address this in this work we binned our objects based on magnitudes.
As mentioned in Section 4.1, a faint end limit on the $r$-band magnitude was placed at 24.5.
Figure 9 of \citet{Balrog_Y3} claims that detection rates of objects at this magnitude were around 20\%, so we decided not to push fainter due to shot noise concerns with such low completion.

%Mention overall goal to have a magnitude bin for areas of bright objects, a transitionary bin, and a bin where detection rates had fallen off significantly.
When binning based on magnitude, general areas of magnitude space with similar properties made sense to be grouped together.
As a general example, bright objects will have nearly perfect detection and classification rates, so that entire region of magnitude space will have very similar corrections.
When detection rates start to fall, this argument no longer applies and finer binning is necessary.
However, with a limited number of \texttt{Balrog} objects to split among the magnitude bins, more bins leads to less statistical significance for calculated rates.
We settled on three bins in an effort to have a magnitude bin for bright objects, a bin for faint objects where detection rates were significantly lower, and a transition bin between those two extremes.

%Describe general test used (average of maximum deviation) to get some idea of performance during Y6 runs.
To determine the magnitude bins used, we focused on how effective they would be in \texttt{Balrog} Y6 where we assumed we would have access to 5 times as many objects.
Since relative detection rates have larger variations than relative classification probabilities, we focused on them for constructing a metric to analyze our magnitude bins.
Once a magnitude bin was chosen and we had the appropriate \texttt{Balrog} objects, various percentage crops would be applied to simulate different injection counts.
We split this smaller sample into an 80\% training sample and a 20\% testing sample.
Training for relative detection rates was performed, and we viewed dependencies from each survey property on relative detection rates within the 20\% testing sample.
Each survey property would have one out of the ten bins as the furthest away from 1.
For this bin we stored the deviation of the relative detection rate from 1 to get an idea of the maximum impact that survey property would have on relative detection rates.
We then averaged this value over all survey properties to get a general idea of the deviations one would expect to be caused by survey properties.
This average value will be referred to as the average maximum deviation.

We found that this average maximum deviation as a function of percentage of \texttt{Balrog} objects used could be modeled well by a power-law function, which we would then extrapolate out to 500\% to get an idea of accuracy expected from Y6.
For relative detection rates, when calculating stellar counts we were only interested in the rates at which stars are classified as stars and galaxies are classified as stars.
Galaxies classified as stars are uncommon at bright magnitudes, so our correction for this relative detection rate is not as consistently important across magnitude bins as our correction for stars classified as stars.
Due to this, we focused on the relative detection rate of stars classified as stars for our work.
Magnitude bins were tested to get similar average maximum deviations for the relative detection rate of stars classified as stars across all three magnitude bins.
When using magnitude bins of $r \leq 22.9$, $22.9<r\leq 23.9$, and $23.9<r\leq 24.5$, we found that the average maximum deviation would take values of 2.2\%, 2.4\%, and 2.5\% respectively when extrapolated to Y6 object counts.
An example of this average maximum deviation as a function of \texttt{Balrog} object percentage used is shown in Figure \ref{fig:bal_num}.
Also shown is a plot generated from detection rates for \texttt{Balrog} galaxies classified as either stars or galaxies.
This was the largest possible sample of objects that could be used for such a test, which made it useful for determining the form of the fitting function as we could probe lower object percentages.

\bibliographystyle{yahapj_twoauthor_arxiv_amp}
\bibliography{main}

% Improved Plots Left: Figure D.1

\end{document}